\documentclass[12pt,a4paper]{article}
\usepackage{type1cm,amsmath,graphicx,indentfirst}
\usepackage{hyperref}
\usepackage[psamsfonts]{amssymb}
\numberwithin{equation}{section}

\DeclareMathOperator{\tr}{tr}

 \newcommand{\ga}{\gamma}
 \newcommand{\be}{\beta}
 \newcommand{\de}{\delta}

 \newcommand{\tha}{\theta}

\newcommand{\half}{\mbox{$\frac{1}{2}$}}

\newcommand{\normord}[1]{{} : #1 : {}}

\newcommand{\del}{\partial}
\newcommand{\delbar}{\bar{\partial}}

\begin{document}
\begin{titlepage}

 \renewcommand{\thefootnote}{\fnsymbol{footnote}}
\begin{flushright}
 \begin{tabular}{l}
 RUP-15-19\\
 \end{tabular}
\end{flushright}

 \vfill
 \begin{center}

\noindent{\large \textbf{Correspondences between WZNW models and CFTs with $W$-algebra symmetry
}}
\vspace{1.5cm}

\noindent{ Thomas Creutzig,$^{a}$\footnote{E-mail: creutzig@ualberta.ca} Yasuaki Hikida$^b$\footnote{E-mail:
hikida@phys-h.keio.ac.jp} and Peter B. R\o nne$^{c,d,e}$\footnote{E-mail: peter.roenne@gmail.com}}
\bigskip

 \vskip .6 truecm
\centerline{\it $^a$University of Alberta,
Department of Mathematical and Statistical Sciences,}
\centerline{\it Edmonton, AB T6G 2G1, Canada}
\medskip
\centerline{\it $^b$Department of Physics, Rikkyo University, }
\centerline{\it  3-34-1 Nishi-Ikebukuro, Toshima, Tokyo 171-8501, Japan}
\medskip
\centerline{\it $^c$ University of Luxembourg, Mathematics Research Unit, FSTC,
} \centerline{\it Campus
Kirchberg, 6, rue Coudenhove-Kalergi, }
\centerline{\it L-1359 Luxembourg-Kirchberg, Luxembourg}
\centerline{\it $^d$ Inria Nancy - Grand Est \& LORIA, France}
\centerline{\it $^e$ SnT, University of Luxembourg, Luxembourg}

 \vskip .4 truecm

 \end{center}

 \vfill
\vskip 0.5 truecm

\begin{abstract}
We study theories with $W$-algebra symmetries and their relation to 
WZNW models on (super-)groups. Correlation functions of the
WZNW models are expressed in terms of correlators of CFTs with $W$-algebra symmetry. The symmetries of the theories involved in these correspondences are related by the Drinfeld-Sokolov reduction of Lie algebras to $W$-algebras. The $W$-algebras considered in this paper are the Bershadsky-Polyakov algebra for sl(3)
and the quasi-superconformal algebra for generic sl$(N|M)$.
The quantum $W$-algebras obtained from affine $\text{sl}(N)$ are constructed using embeddings of sl$(2)$ into sl$(N)$, and these can in turn be characterized by partitions of $N$. The above cases correspond to $\underline{N+2} = \underline{2} + N \underline{1}$ and its supergroup extension. Finally, sl$(2N)$ and the correspondence corresponding to
$\underline{2N} = N \underline{2}$ is also analyzed.
\end{abstract}
\vfill
\vskip 0.5 truecm

\setcounter{footnote}{0}
\renewcommand{\thefootnote}{\arabic{footnote}}
\end{titlepage}

\newpage

\tableofcontents

\section{Introduction}

Two-dimensional conformal field theories play an important role in both theoretical physics and mathematics. Their infinite-dimensional symmetries restrict the theories to a large extent, but leave enough room for interesting structures and intriguing dualities.

When the basic Virasoro symmetry is complemented with 
higher spin generators up to spin $N$, it is known as $W_N$-algebra symmetry, see \cite{Bouwknegt:1992wg} for a nice review.
Recently, theories with $W$-symmetry have appeared in several contexts.
A prominent example is the AdS/CFT correspondence.
It was shown in \cite{Henneaux:2010xg,Campoleoni:2010zq} that the asymptotic symmetry of higher spin gravity theory on AdS$_3$ can be
identified with $W$-symmetry. Based on this fact, the authors of \cite{Gaberdiel:2010pz} proposed
that the minimal model with $W_N$-algebra symmetry in a large $N$ limit is dual to the higher spin gravity
by Prokushkin and Vasiliev \cite{Prokushkin:1998bq}.
The higher spin $W$-symmetry plays an important role in the evidence of this
AdS/CFT correspondence.
Supersymmetric versions of the higher spin AdS/CFT correspondence were also proposed, e.g., in \cite{Creutzig:2011fe,Candu:2012jq,Henneaux:2012ny,Creutzig:2012ar,Beccaria:2013wqa,Gaberdiel:2013vva,Creutzig:2013tja,Creutzig:2014ula,Gaberdiel:2014cha}, where super $W$-algebras appear as symmetry algebras.
The theories with $W_N$-symmetry also appear as effective descriptions of subsectors of
four-dimensional SU$(N)$ gauge theories \cite{Alday:2009aq,Wyllard:2009hg}.

In this paper, we study some aspects of two-dimensional conformal field theories with $W$-symmetry.
More concretely, we relate correlation functions of
Wess-Zumino-Novikov-Witten (WZNW) models on the groups SL$(3)$ and SL$(2N)$ to correlators of theories with $W$-symmetry, and similarly we consider the correspondence between WZNW models on the supergroups SL$(N|M)$ and theories with super $W$-algebra symmetry. Already in \cite{Ribault:2005wp} Ribault and Teschner showed explicitly that there is a relation between $N$-point amplitudes of primary operators on spheres in the
$H_3^+$ WZNW model, which describes strings in Euclidean AdS$_3$, and $(2N-2)$-point spherical amplitudes in Liouville field theory.  In \cite{HS} this relation was re-derived using an
intuitive path integral method and extended the correspondence to amplitudes on Riemann surfaces of arbitrary genus.
Using this method further generalizations were possible:
The relation between correlation functions of the OSP$(p|2)$ WZNW model and of ${\cal N}=p$
supersymmetric Liouville field theory was obtained in \cite{HS2} for $p=1,2$.
In our previous paper \cite{CHR}, we have extended these relations
to the cases with WZNW models on supergroups whose bosonic subgroup
is of the form SL$(2) \times A$.  For example, the PSU$(1,1|2)$ WZNW model is related to the
small ${\cal N}=4$ super Liouville field theory, and the OSP$(p|2)$ WZNW model with $p \geq 3$
is related to a superconformal field theory with SO$(p)$-extended super conformal symmetry
discussed in \cite{K,B}.
Other examples considered in that paper are with the supergroups  SL$(2|p)$,
D$(2,1;\alpha)$, OSP$(4|2p)$, F$(4)$ and G$(3)$, see also \cite{KW}. Especially the case of D$(2,1;\alpha)$ relates to the large ${\cal N}=4$ super conformal algebra.
This paper is a continuation of these works.

It is known that quantum $W$-algebras can be obtained by quantum Drinfeld-Sokolov
reduction of affine Lie algebras, which are the symmetries of WZNW models \cite{Bouwknegt:1992wg}.
We think of the relations between correlation functions mentioned above as
correlator versions of Drinfeld-Sokolov reduction.
There are different
quantum $W$-algebras associated to a given simple Lie algebra corresponding to different embeddings of a sl(2) subalgebra into the Lie algebra, see e.g. \cite{deBoer:1993iz}.
The principal embedding leads to the $W_N$-algebra starting from the affine sl$(N)$ algebra; but we are interested in other embeddings.
The simplest non-trivial case appears for  sl(3), where the non-principal embedding leads to the Bershadsky-Polyakov algebra
\cite{Polyakov,Bershadsky}. For sl$(N)$, the sl(2) embedding can be determined by the
branching of the fundamental representation $\underline{N}$, which can be expressed by
a partition of $N$ (see, e.g., \cite{Bais:1990bs} in this context). The principal embedding
corresponds to $\underline{N} = \underline{N}$.
For sl(3), the principal embedding $\underline{3} = \underline{3}$ leads to $W_3$ algebra
firstly introduced in \cite{Zamolodchikov:1985wn},
and the Bershadsky-Polyakov algebra
comes from the embedding $\underline{3} = \underline{2} + \underline{1}$.
We also consider the embedding $\underline{N+2} = \underline{2} + N \underline{1}$,
which leads to a generalized Bershadsky-Polyakov algebra  also called the quasi-superconformal algebra \cite{Romans}. The algebra is like the Knizhnik-Bershadsky
algebra \cite{K,B}, and we note that they all have bosonic spin-3/2 fields.
Further, the non-trivial embedding given by $\underline{2N} = N\underline{2}$ is also discussed.
Furthermore, supergroup cases are examined.

Once these relations are established, there are many ways to utilize them.
In order to investigate the AdS/CFT correspondence, it is important to analyze
superstrings on AdS spaces, which may be described by supergroup WZNW models.
One of the applications is thus to use the relations for the study of supergroup models.
In fact, some structure constants of OSP$(1|2)$ models are computed in terms of
${\cal N}=1$ super Liouville field theory in \cite{HS2,Creutzig:2010zp}.
Another important use of this type of relation was in a proof of the
Fateev-Zamolodchikov-Zamolodchikov (FZZ) duality \cite{FZZ,HS3,Creutzig:2010bt}.
The FZZ conjecture can be viewed as a T-duality in curved space, and the T-dual of the two-dimensional Euclidean black hole is proposed to be sine-Liouville theory.
The black hole model can be described by the SL(2)/U(1) coset, and the application of the
Ribault-Teschner relation to the SL(2) part is an essential part of the proof in \cite{HS3}.
Our hope is that similar relations naturally lead to generalizations  of the FZZ dualities.
We will comment on this in the conclusion.

The organization of this paper is as follows.
In the next section, we derive a relation between correlation functions of the SL$(3)$ WZNW model
and of a theory with Bershadsky-Polyakov symmetry.
In section \ref{QSCA}, we extend the relation to the more general case where the
SL$(N+2|M)$ WZNW model is related to a theory with quasi-superconformal symmetry.
In section \ref{product}, we study the theory obtained by the Drinfeld-Sokolov reduction of the SL$(2N)$
WZNW model with the product embedding $\underline{2N}=N\underline{2}$.
In section \ref{extention}, we make a proposal of how to combine the cases analyzed so far.
Some technical details are collected in appendices. In appendix \ref{conv}, the conventions
for the Lie (super-)algebras are summarized. In appendix \ref{genus}, we extend the
analysis in section \ref{QSCA} to the cases with Riemann surfaces of arbitrary genus.
In appendix \ref{DSreduction}, our correspondence is compared with the known facts
on the Hamiltonian reduction with the product embedding $\underline{2N}=N\underline{2}$.
In appendix \ref{non-hol}, some field redefinitions used in section \ref{product} are derived.

\section{Bershadsky-Polyakov algebra and SL(3)}

We start our discussion with the interesting example of a theory with the Bershadsky-Polyakov algebra as symmetry algebra.
This algebra is a Drinfeld-Sokolov reduction of $\widehat{\text{sl}}(3)$, and
the Bershadsky-Polyakov algebra corresponds to
the non-principal embedding of $\text{sl}(2)$ in $\text{sl}(3)$.
Conventions on $\text{sl}(3)$ and the non-principal embedding are provided in Appendix \ref{subsec:conv}.

Given this embedding, the $\text{sl}(2)$ subalgebra acts on $\text{sl}(3)$. Especially, we can decompose $\text{sl}(3)$ into eigenspaces of the Cartan subalgebra of $\text{sl}(2)$.
This leads to a $\mathbb Z_5$-gradation
\begin{equation}
\text{sl}(3) \ = \ \mathfrak{g}_{-1}\oplus\mathfrak{g}_{-1/2}\oplus\mathfrak{g}_{0}\oplus\mathfrak{g}_{1/2}\oplus\mathfrak{g}_{1}.
\end{equation}
This gradation is the starting point for the construction of the map between correlation functions in the SL(3) WZNW-theory and a theory with the Bershadsky-Polyakov algebra as symmetry algebra.

\subsection{SL(3) WZNW action}

In order to construct the action of the WZNW theory, we start with a SL(3)-valued field $g$. For our purposes it is important to parameterize this field
according to above $\mathbb Z_5$-gradation, namely
\begin{align}
 g=g_{-1}\,g_{-\frac{1}{2}}\,g_{0}\, g_{\frac{1}{2}}\,g_{1}=e^{\ga_{-1}}e^{\ga_{-1/2}}e^{\phi_0}e^{\ga_{1/2}}e^{\gamma_1} ~.
\end{align}
This is similar to the case of SL$(2|1)$ studied in \cite{HS2,CHR}, but of course $\gamma_{\pm 1/2}$ are here bosonic fields.

The action of the SL(3) WZNW model at level $k$ is
\begin{align}
 S^\text{WZNW} [g] = \frac{k}{4 \pi} \int_{\Sigma} d ^2z
 \langle g^{-1} \partial g , g^{-1} \bar \partial g \rangle
 + \frac{k}{2 4 \pi} \int_{B} \langle g^{-1} d g ,
  [g^{-1} d g , g^{-1} d g ] \rangle
\end{align}
with $\partial B = \Sigma$.
The invariant bilinear form is of course the Killing form.
The Polyakov-Wiegmann identity
\begin{align}
\label{PWid}
 S^\text{WZNW}  [ gh ]  =  S^\text{WZNW} [g] +  S^\text{WZNW} [h]
 + \frac{k}{2 \pi} \int d^2 z
   \langle g^{-1} \bar \partial g , \partial h h^{-1} \rangle
\end{align}
leads to
\begin{equation}
 \begin{split}
  S^\text{WZNW} [g]
   =& S^\text{WZNW} [g_{0}]
     + \frac{k}{2\pi} \int d^2 z \langle \big(\delbar \ga_{-1}-\frac{1}{2}[\gamma_{-1/2},\delbar\ga_{-1/2}]+
     \delbar\gamma_{-1/2}\big), \\
     &\qquad\qquad \qquad \qquad \mathrm{Ad}(g_0)\big(\del \ga_{1}+\frac{1}{2}[\gamma_{1/2},\del\ga_{1/2}]+\del\gamma_{1/2}\big)   \rangle ~.
\end{split}
\end{equation}
Here, we used that the Killing form respects the $\mathbb Z_5$-gradation. We can now introduce the auxiliary fields $\be_{1},\be_{-1}$ with $\be_{1}\in{g}_{1},\be_{-1}\in{g}_{-1}$ such that the integration over
$\be_{1},\be_{-1}$  reproduces the original action.
Likewise, we also introduce auxiliary variables for the half-integer parts, $\be_{1/2}, \be_{-1/2}$ with $\be_{1/2}\in{g}_{1/2},\be_{-1/2}\in{g}_{-1/2}$. The action now takes the form
\begin{equation}
 \begin{split}
  S^\text{WZNW} [g]&=S^\text{WZNW}_{\textrm{ren}} [g_{0}] + S_0+S_{\textrm{int}} ~,  \\
   S_0 &= \frac{k}{2\pi} \int d^2 z
   \Big[ \langle \be_{1}, \delbar\gamma_{-1} \rangle  + \langle \be_{1/2}, \delbar\gamma_{-1/2}\rangle + \langle \be_{-1},
   \del\gamma_{1} \rangle  + \langle \be_{-1/2}, \del\gamma_{1/2}  \rangle \Big] ~ , \\
    S_{\textrm{int}} &= -\frac{k}{2\pi} \int d^2 z  \Big [  \langle \be_{1},\ \mathrm{Ad}(g_0)( \be_{-1})   \rangle \\
     &\qquad\qquad\qquad+ \langle \big(\be_{1/2}-\frac{1}{2}[\gamma_{-1/2},\be_{1}]\big), \mathrm{Ad}(g_0)\big( \be_{-1/2}+\frac{1}{2}[\gamma_{1/2},\be_{-1}]\big)\rangle \Big]~,
\end{split}
\end{equation}
where we have used the invariance of the inner product.
The quantum action $S^\text{WZNW}_{\textrm{ren}} (g_{0})$ is then obtained
by taking care of the Jacobian which appears due to the introduction of the new auxiliary fields.

This is as far we can go using just the knowledge of the 5-gradation. We can now insert the SL(3) generators
(see appendix \ref{subsec:conv}), we define
\begin{align}
g=e^{\gamma^3 F_3}e^{\gamma^1 F_1+\gamma^2 F_2}e^{\phi\theta+\phi^\bot\theta^\bot}e^{\bar \gamma^1 E_1+\bar\gamma^2 E_2}e^{\bar\gamma^3 E_3}\, ,
\end{align}
and correspondingly (abusing notation -- on the left hand side the fields are the old Lie algebra valued fields, and the $\be_i$s on the right hand side are the new fields).
\begin{equation}
 \begin{split}
 \be_{1} &= \be_3 E_3\, ,\qquad\qquad\quad\,\qquad  \be_{-1} = \bar\be_3 F_3\, , \\
  \be_{1/2} &=\be_1 E_1+\be_2 E_2\, ,\qquad\,\qquad  \be_{-1/2} =\bar \be_1 F_1+\bar \be_2 F_2\, .
\end{split}
\end{equation}
The action then takes the form
\begin{equation}
 \begin{split}
  S^\text{WZNW} [g]&=S^\text{WZNW}_{\textrm{ren}} [g_{0}] + S_0+S_{\textrm{int}} ~, \\
   S^\text{WZNW}_{\textrm{ren}}  [g_{0} ] &=\frac{1}{4\pi} \int d^2 z \Big[\frac{1}{2}\del\phi\delbar\phi+\frac{3}{2}
   \del\phi^\bot\delbar\phi^\bot-\frac{1}{2}b\sqrt{g}\mathcal{R}\phi \Big]  ~,  \\
   S_0 &= \frac{1}{2\pi} \int d^2 z \Big[ \be_{1} \delbar\gamma^{1} +  \be_{2} \delbar\gamma^{2}+  \be_{3}
   \delbar\gamma^{3}+ \bar\be_1 \del\bar\gamma^{1}+ \bar\be_2 \del\bar\gamma^{2}+ \bar\be_3 \del\bar\gamma^{3} \Big] ~,  \\
    S_{\textrm{int}} &= -\frac{k}{2\pi} \int d^2 z  \Big [  \be_3\bar\be_3 e^{-b\phi}  + (\be_1+\frac{1}{2}\be_3\ga^2)
    (\bar\be_1+\frac{1}{2}\bar\be_3\bar\ga^2)e^{-b\phi/2-3b\phi^\bot/2}  \\
    &\qquad\qquad \qquad +(\be_2-\frac{1}{2}\be_3\ga^1)(\bar\be_2-\frac{1}{2}\bar\be_3\bar\ga^1) e^{-b\phi/2+3b\phi^\bot/2}\Big]~,
\end{split}
\end{equation}
where we have rescaled the fields $\phi,\phi^\bot$ with $b=1/\sqrt{k-3}$. The equations of motion for the auxiliary fields take the form (before this rescaling)
\begin{align}
 \be_3 &= ke^{\phi}\big(\del\bar\gamma_3+\frac{1}{2}\bar\ga^1\del\ga^2-\frac{1}{2}\bar\ga^2\del\ga^1\big)\, , \\
 \be_2 &= ke^{\phi/2-3\phi^\bot/2}\del\bar\gamma^2+\frac{1}{2}\gamma^1 \beta_3\, ,\\
 \be_1 &= ke^{\phi/2+3\phi^\bot/2}\del\bar\gamma^1-\frac{1}{2}\gamma^2 \beta_3\, .
\end{align}
This tells us the correct renormalization of the fields $\phi,\phi^\bot$ for the action $S^\text{WZNW}_{\textrm{ren}}$.
Note that the background charge implies that the first interaction term $\be_3\bar\be_3 e^{-\phi}$ is not marginal,
but it can be seen as a contact term and neglected. The sl(3) currents can be written in terms of these free fields
as is demonstrated in appendix \ref{free}.

\subsection{A relation to Bershadsky-Polyakov algebra}
\label{BPalgebra}

We would now like to derive the relation between correlation functions,
so we begin by considering a correlator in the SL(3) WZNW model.
The vertex operators can always be chosen of the form
\begin{align}\label{}
V_\nu(z_\nu)=e^{\mu_3^\nu\gamma^3-\bar\mu_3^\nu\bar\gamma^3}f(\gamma^1,\ga^2,\bar\gamma^1,\bar\ga^2,\phi,\phi^\bot) \, ,
\label{sl3vertexop}
\end{align}
and we consider correlators
\begin{align}
 \langle \prod _{\nu=1}^n V_\nu (z_\nu) \rangle
  &=  \int  {\cal D} g  \, e^{-S^\text{WZNW} (g)} \prod_{\nu=1}^{n} V_\nu ( z _ \nu) ~,
  &{\cal D} g  &=
 {\cal D} \phi {\cal D} \phi^\bot  \prod_{\alpha =1 }^3
 {\cal D}^2 \beta_\alpha {\cal D}^2 \gamma^\alpha ~.
 \label{correlator}
\end{align}
Later we fix the form of the functional $f(\gamma^1,\ga^2,\bar\gamma^1,\bar\ga^2,\phi,\phi^\bot)$, but
here we leave it arbitrary.

Following \cite{HS} we integrate out $\gamma^3$ which only appears linearly in exponents. This gives a delta function in $\beta_3$ which is solved by setting
\begin{align}\label{}
      \be_3(z)\mapsto {\cal B}_3(z)=\sum_{\nu=1}^n \frac{\mu^\nu_3}{z - z_\nu}
  = u \frac{\prod_{i=1}^{n-2} (z - y_{i})}
             {\prod_{\nu=1}^{n} (z - z_\nu)}\, .
\end{align}
Similarly for the antiholomorphic side we have
\begin{align}\label{}
      \bar\be_3(\bar z)\mapsto -\mathcal{\bar B}_3(\bar z)=-\sum_{\nu=1}^n \frac{\bar\mu^\nu_3}{\bar z - \bar z_\nu}
  = -\bar u \frac{\prod_{i=1}^{n-2} (\bar z - \bar y_{i})}
             {\prod_{\nu=1}^{n} (\bar z - \bar z_\nu)}\, .
\end{align}
We then remove the function $\mathcal{B}_3$ from the action by the following transformation:
\begin{align}\label{eq:changetohalfhalf}
    {\gamma'}^a= ( u\mathcal{B}_3 )^{\half} \gamma^a ~, \qquad   \beta'_a&= ( u \mathcal{B}_3)^{-\half}\beta_a ~, \qquad
    \varphi =\phi-\frac{1}{2b} \ln |u \mathcal{B}_3 |^2~ , \qquad \varphi^\bot =\phi^\bot ~.
\end{align}
This will change the background charge of $\varphi$, and
the remaining $\beta'_1,\gamma'^1$, $\beta'_2,\gamma'^2$ ghost systems are now dimension $(1/2,1/2)$ systems. This also gives extra terms of the form
\begin{align}\label{eq:extraghostterms}
    \de S=\frac{1}{2\pi} \int d^2 z \Big[ -\half\be'_{i}\gamma'^{i}\delbar\ln\mathcal{B}_3 -\half\bar\be'_{i}\bar\gamma'^{i}\del\ln\mathcal{\bar B}_3\Big]\, .
\end{align}
A natural way to proceed is to use that $\del\delbar\ln\mathcal{B}_3$ can be expressed by delta functions localized at the points $z_\nu$ and $y_i$.
The extra terms can thus put in the form of vertex operators if we bosonize the $\be,\ga$ systems as follows
\begin{align}\label{eq:simplebosonization}
 \be'_1 =  \partial Y_1 e^{- X_1 +  Y_1}~, \qquad
 \ga'^1 = e^{X_1 -  Y_1}  ~, \qquad \normord{\be'_1\ga'^1}=\del X_1\, ,
\end{align}
and likewise for $\be'_2,\ga'^2$.
We then get extra insertions in the correlators depending on $X_1,X_2$ and $\varphi$ and these are located in the zeroes of $\mathcal{B}_3$, namely $y_{i}$. Further, the original vertex operators in $z_\nu$ also gets modified.
An explicit formula can be found in equation \eqref{eq:CorrRela} in the next section.

On the other hand, we can also do a SP(2) transformation of the $\be',\ga'$ systems as follows
\begin{align}
 \be'_1=\half(\ga-\ga_d)\, , \qquad \ga'^1=-\be+\be_d\, , \qquad
 \be'_2=\half(\be+\be_d)\, , \qquad \ga'^2=\ga+\ga_d\, ,\label{eq:sp2trans}
\end{align}
and similarly for the anti-holomorphic side
\begin{align}
 \bar\be'_1=\half(\bar\ga-\bar\ga_d)\, , \qquad \bar\ga'^1=-\bar\be+\bar\be_d\, ,
 \qquad
 \bar\be'_2=-\half(\bar\be+\bar\be_d)\, , \qquad \bar\ga'^2=-\bar\ga-\bar\ga_d\, .
\end{align}
The point is that this decouples the $\beta_d,\ga_d$ system and, as we will see below, the remaining action has Bershadsky-Polyakov symmetry. After the transformation, the action takes the form
\begin{equation}
 \begin{split}
  &S=S_{\textrm{kin}} + S_{\textrm{int}} ~,  \\
  &S_{\textrm{kin}}  =\frac{1}{4\pi} \int d^2 z \Big[\frac{1}{2}\del\varphi\delbar\varphi+\frac{3}{2}
  \del\varphi^\bot\delbar\varphi^\bot-\frac{1}{2}(b+\frac{1}{2b})\sqrt{g}\mathcal{R}\varphi    \\
	&\qquad\qquad\qquad\qquad +2\be \delbar\gamma+2\be_{d} \delbar\gamma_{d} + 2\bar\be \del\bar\gamma +
	2\bar\be_d \del\bar\gamma^{d} \Big] ~, \\
    &S_{\textrm{int}} = -\frac{k}{2\pi} \int d^2 z  \Big [   \ga\bar\ga e^{-b\varphi/2-3b\varphi^\bot/2}  -\be\bar\be e^{-b\varphi/2+3b\varphi^\bot/2}\Big]~.
\end{split}
\end{equation}
We have here omitted the contact term in the action.
However, the holomorphic part of the extra ghost terms \eqref{eq:extraghostterms} are now
\begin{align}
    \de S=\frac{1}{2\pi} \int d^2 z \Big[ -\half(\ga\be_d+\ga_d\be)\delbar\ln\mathcal{B}_3 -\half(\bar\ga\bar\be_d+\bar\ga_d\bar\be)\del\ln\mathcal{\bar B}_3 \Big]\, .
\end{align}
Since the logs are holomorphic up to branch cuts, these terms look more like line operators than the extra vertex operators  that were found after the bosonization \eqref{eq:simplebosonization} above. To perform the SP(2) rotation on the vertex operators is thus not a simple problem, however in the similar OSP$(N|2)$ case \cite{HS3,CHR} the corresponding problem for the fermionic ghost systems was solvable using spin operators. We are thus hopeful that future research will find a solution. 

Let us now consider the symmetry of the action. We want to show that the action with the $\be_d,\ga^d$ part  removed is invariant under the Bershadsky-Polyakov $W_3^2$ algebra \cite{Bershadsky}. This algebra is generated by dimension $3/2$ fields $G^\pm$ and the U(1) current $H$ with the OPEs
\begin{align}
 &G^+(z)G^-(w)\sim \frac{(k-1)(2k-3)}{(z-w)^3}+\frac{-3(k-1)H(w)}{(z-w)^2} \\
 & \qquad \qquad \qquad \qquad +\frac{\big(3:HH:+(k-3)T_{W_3^2}-\frac{3(k-1)}{2}\del H\big)(w)}{z-w} ~, \nonumber\\
&G^\pm(z)G^\pm(w)\sim 0 ~, \qquad
H(z)H(w)\sim\frac{-(2k-3)/3}{(z-w)^2} ~. \nonumber
\end{align}
The task is to construct these currents in terms of our free fields.

We have removed $\mathcal{B}_3$ from the action, and
it changes the stress energy tensor to
\begin{align}
 T_{\textrm{improved}}=T_{sl_3}-\del J^\tha ~,
\end{align}
which is called the improved stress-energy tensor in \cite{Bershadsky}. We here have to think of the $\be_3,\ga^3$ systems as being removed. After doing the SP(2) transformation \eqref{eq:sp2trans} we get
the stress-energy tensor
\begin{align}
  &T=T_{W_3^2}+T_{\be_d,\ga_d} ~ , \nonumber\\
  &T_{W_3^2} = -\tfrac{1}{4}\del\varphi\del\varphi-\tfrac{3}{4}\del\varphi^\bot\del\varphi^\bot -(b+\frac{1}{2b})\del\del\varphi +\half(\ga\del\be-\del\ga\be ) ~ , \nonumber \\
  &T_{\be_d,\ga_d} =\half(\ga_d\del\be_d-\del\ga_d\be_d ) ~ .
\end{align}
Here $T_{W_3^2}$ exactly has the central charge $25+24/(k-3)+6(k-3)$ as expected for the level $k$ Bershadsky-Polyakov algebra.
The $G^\pm,H$ currents are obtained from the currents where first all terms depending on $\ga^3$ is removed, then $\mathcal{B}_3$
is taken out, and finally, after the SP(2) transformation, the terms depending on $\be_d,\ga_d$ are removed. Then we get (the currents are like in \cite{deBoer:1992sy}):
\begin{align}
& G^+=J^{E_2}|_{\textrm{reduced}} = \frac{1}{2b}\del\varphi\ga-\frac{3}{2b}\del\varphi^\bot\ga-\ga\ga\be +(k-1)\del\ga ~, \nonumber \\
& G^-=J^{E_1}|_{\textrm{reduced}} =-\frac{1}{2b}\del\varphi\be-\frac{3}{2b}\del\varphi^\bot\be-\ga\be\be -(k-1)\del\be ~,  \\
& H=-\frac{2}{3}J^{\tha^\bot}|_{\textrm{reduced}} =-\frac{1}{b}\del\varphi^\bot -\ga\be ~.\nonumber
\end{align}
Finally, our interaction terms are screening operators for the $W_3^2$ algebra which can be directly checked, and we have thus demonstrated that the action is invariant under the Bershadsky-Polyakov algebra.

\section{Quasi-superconformal theory from $\text{SL}(2+N|M)$}
\label{QSCA}

In this section we generalize the analysis for sl(3) in the previous section to
the case with a supergroup  $\text{SL}(2+N|M)$ with arbitrary numbers $N,M$.
In \cite{CHR}, we have derived relations between correlation functions
of supergroup WZNW models and extended super Liouville theory.
The supergroup used in \cite{CHR} has a bosonic subalgebra of the
form $\text{SL}(2) \times A$. The procedure developed in \cite{HS} applied to the SL(2) part led to the
correspondence found in \cite{CHR}. Here we use a similar embedding
of sl(2) such that the elements of the Lie superalgebra are
decomposed as
\begin{align} \label{5dec}
 \mathfrak{g}=\mathfrak{g}_{-1} \oplus \mathfrak{g}_{-\frac{1}{2}} \oplus \mathfrak{g}_{0} \oplus \mathfrak{g}_{+\frac{1}{2}} \oplus \mathfrak{g}_{+1} ~.
\end{align}
The generators of the embedded sl(2) are in $t^z \in \mathfrak{g}_0 $ and
$t^\pm \in \mathfrak{g}_{\pm 1}$.
In fact, $\mathfrak{g}_{\pm 1}$ are generated by $t^\pm$.
Notice that $\mathfrak{g}_{\pm 1/2}$
are fermionic in the cases considered in \cite{CHR}, but here they can be bosonic
and fermionic. This is the main point of the generalization.

The relations in \cite{HS,HS2,CHR} remind us of Hamiltonian reduction
\cite{Bouwknegt:1992wg},
and one of the aims of this paper is to investigate the relation between the two.
We can construct a theory with $W$ symmetry from the WZNW model on $G$,
and the symmetry algebra depends on how sl(2) is embedded in $g$.
The above sl(2) embedding  corresponds to the partition of $2+N|M$ as
\begin{equation}
\underline{2+N|M}  = \underline{2|0}+N \underline{1|0}+M \underline{0|1}\, .
\end{equation}
For $N=0$ the supergroup is given by SL$(2|M)$ and hence analyzed in \cite{CHR}, and
for $M=0$ the Hamiltonian reduction of sl$(2+N)$ yields the quasi-superconformal algebra
in \cite{Romans}. Here we deal with generic $N,M$, which involves a generalization of the
quasi-superconformal algebra.

\subsection{SL$(2+N|M)$ WZNW action}
\label{lotone}

The elements of the sl$(2+N|M)$ Lie superalgebra
can be expressed by the $(2+N|M) \times (2+N|M)$ supermatrix of the form
\begin{align} M =
 \begin{pmatrix}
   A & B \\
   C & D
  \end{pmatrix} ~, \qquad
  \text{str} \, M = \text{tr} \, A - \text{tr} \, D = 0 ~,
\end{align}
where $A,D$ are Grassmann even and $B,C$ are Grassmann odd.
Following the general argument, we decompose sl$(2+N|M)$ as in \eqref{5dec}.
Roughly speaking, we interpret the  $(2+N|M) \times (2+N|M)$ supermatrix
as a four block supermatrix, where the diagonal blocks are a $2\times 2$ block representing the embedded sl(2) and
the $(N|M) \times (N|M)$ block representing a sl$(N|M)$ algebra which commutes with the sl(2) and
further there is also a u(1) algebra.
The off-diagonal
part carries the standard representation of sl$(N|M)\oplus$ sl(2) and its conjugate, and they will represented by free bosons and fermions. See for instance
\cite{Bars}.
For our explicit conventions see appendix \ref{convgen}.
We parameterize a supergroup valued field according to our 5-decomposition
\begin{align}
 g=g_{-1}\,g_{-\frac{1}{2}}\,g_{0}\, g_{\frac{1}{2}}\,g_{1} ~,
\end{align}
where
\begin{align}
 g_{-1} = e^{\gamma E^-} ~, \qquad g_{+1} = e^{\bar \gamma E^+} ~,
 \qquad
 g_0 =  e^{ - 2 \Phi Q} e^{ - 2 \phi E^0 }
\begin{pmatrix}
  \mathbb{I}_2 & 0 \\
  0 & q
\end{pmatrix} ~ .
\end{align}
The generators of the embedded sl(2) are $E^0,E^\pm$ and the elements
of sl$(N|M)$ are represented by $q$. The generator of the u(1) algebra
is denoted by $Q$, and it commutes with the sl(2) and sl$(N|M)$.
For $2+N=M$, the u(1) part can be decoupled and we can start
from psl$(M|M)$ instead of sl$(M|M)$. The other parts are
\begin{align}
& g_{-1/2} = \exp (\sum_{i=1}^N \gamma^1_i S^-_{1,i} )
          \exp (\sum_{i=1}^N \gamma^2_i S^-_{2,i} )
          \exp (\sum_{\hat i=1}^M \theta^1_{\hat i} F^-_{ 1,\hat i} )
          \exp (\sum_{\hat i=1}^M \theta^2_{\hat i} F^-_{2,\hat i} ) ~, \\
& g_{+1/2} = \exp (\sum_{\hat i=1}^M \bar \theta^2_{\hat i} F^+_{2,\hat i} )
         \exp (\sum_{\hat i=1}^M \bar \theta^1_{\hat i} F^+_{1,\hat i} )
         \exp (\sum_{i=1}^N \bar \gamma^2_i S^+_{2,i} )
         \exp (\sum_{i=1}^N \bar \gamma^1_i S^+_{1,i} ) ~.
\end{align}
In \cite{CHR} we have only fermions $\theta^a_i, \bar \theta^a_i$, and
the appearance of bosons $\gamma^a_i , \bar \gamma^a_i$ is the new
feature in this case. The explicit form of the generators are summarized in
appendix \ref{convgen}.

With the above parametrization, the action of SL$(2+N|M)$ WZNW model
is given by
\begin{align}
 S^\text{WZNW}_k[g] = S^\text{WZNW}_k[q] &+ \frac{k}{2 \pi}
  \int d z^2 \Bigl[ \bar \partial \Phi \partial \Phi
 + \bar \partial \phi \partial \phi   + e^{-2 \phi}
 (\bar \partial \gamma + \partial \Theta_1  \Theta_2^t )
  (\partial \bar \gamma + \bar \Theta_2 \partial \Theta_1^t ) \nonumber
\\
& + e^{- \phi + \eta \Phi}
 \partial \bar \Theta_1 q^{-1} \bar \partial \Theta_1^t
 +  e^{- \phi - \eta \Phi }  \bar \partial \Theta_2 q
 \partial  \bar \Theta_2^t  \Bigr ] ~,
\end{align}
where
\begin{align}
\nonumber
  \gamma_a = (\gamma_1^a , \cdots , \gamma_N^a) ~,
\quad \bar \gamma_a = (\bar \gamma_1^a , \cdots , \bar \gamma_N^a)  ~, \quad
  \theta_a = (\theta_1^a , \cdots , \theta_M^a) ~,
\quad \bar \theta_a = (\bar \theta_1^a  , \cdots , \bar \theta_M^a)
\end{align}
with $a=1,2$.  Moreover we denote $\Theta_a = (\gamma_a , \theta_a)$
and $\bar \Theta_a = (\bar \gamma_a , \bar \theta_a)$.
We introduce auxiliary fields $\beta, \bar \beta,
P_a = (\beta_a , p_a), \bar P_a = (\bar \beta_a , \bar p_a) $ with
\begin{align}
\nonumber
\beta_a=(\beta^a_1 , \cdots , \beta^a_N ) ~, \quad
\bar \beta_a=(\bar \beta^a_1 , \cdots , \bar \beta^a_N ) ~, \quad
p_a=(p^a_1 , \cdots , p^a_M ) ~, \quad
\bar p_a=(\bar p^a_1 , \cdots , \bar p^a_M ) ~.
\end{align}
Then we find classically
\begin{align}
 S^\text{WZNW}_k[g] &\stackrel{\text{clas.}}{=} S^\text{WZNW}_k [q]+ \frac{1}{2 \pi}
 \int d^2 z \Bigl[ k \bar \partial \Phi \partial \Phi +
 k \bar \partial \phi \partial \phi -
  \beta \bar \partial \gamma - \bar \beta \partial \bar \gamma
  + \sum_{a=1}^2 ( P_a \bar \partial \Theta_a^t
  + \bar P_a \partial \bar \Theta_a^t  )  \nonumber \\
 & - \frac{1}{k} \beta \bar \beta e^{2 \phi}
  - \frac{1}{k} (P_1 + \beta \zeta \Theta_2) q \zeta
  (\bar P_1 + \bar \beta \bar \Theta_2)^t e^{\phi - \eta \Phi}
 - \frac{1}{k} \bar P_2  q^{-1} \zeta  P_2^t e^{\phi + \eta \Phi} \Bigr]~,
\end{align}
where we have used $P_a \partial \Theta_a^t
= \partial \Theta_a \zeta P_a^t$ with
\begin{align}
\zeta =
 \begin{pmatrix}
   \mathbb{I}_N & 0 \\
  0 & - \mathbb{I}_{M}
 \end{pmatrix} ~.
\end{align}

Due to the anomaly from the change of path integral measure there are shifts of coefficients. To get these corrections, we first set $q=1$.
Then  from the measure of the path integral the contribution
from $\beta,\gamma$ is
\begin{align}
 - \frac{1}{\pi} \int d^2 z \partial \phi \bar \partial \phi
  + \frac{1}{8 \pi} \int d^2z  \sqrt{g} {\cal R } \phi ~.
\end{align}
Further, let
\begin{align}
 \delta s =- \frac{1}{4 \pi} \int d^2 z \partial \phi \bar \partial \phi
  + \frac{1}{16 \pi} \int d^2z  \sqrt{g} {\cal R } \phi ~,
\end{align}
then a pair of $\beta_a,\gamma_a$ contributes $\delta s$, while a pair of $p_a,\theta_a$ contributes $-\delta s$ to the anomaly (see (2.22) of \cite{CHR}).
Taking this anomaly into account, the action becomes
\begin{align} \label{actionslnm}
 & S^\text{WZNW}_k[g] = S^\text{WZNW}_{k-2}[q]  \\ & \qquad + \frac{1}{2 \pi}
 \int d^2 z \Bigl[ \bar \partial \Phi \partial \Phi +
 \bar \partial \phi \partial \phi
   + \frac{\hat Q}{4}  \sqrt{g} {\cal R } \phi
- \beta \bar \partial \gamma - \bar \beta \partial \bar \gamma
  + \sum_{a=1}^2 ( P_a \bar \partial \Theta_a^t
  + \bar P_a \partial \bar \Theta_a^t  ) \nonumber  \\
 & \qquad  - \frac{1}{k} \beta \bar \beta e^{2 b \phi}
  - \frac{1}{k} (P_1 + \beta \zeta \Theta_2) q \zeta
  (\bar P_1 + \bar \beta \bar \Theta_2)^t e^{b (\phi - \eta \Phi)}
 - \frac{1}{k} \bar P_2 q^{-1} \zeta P_2^t e^{b(\phi + \eta \Phi)} \Bigr] \nonumber
\end{align}
with $b^{-2} = k - 2 - N + M$ and $\hat Q = b (1 + N - M)$.
The central charge of the SL$(2+N|M)$
WZNW model is
\begin{align}
 c = \frac{((N-M+2)^2 - 1) k}{k -2 - N+M} ~.
\end{align}
After the renormalization, we have
\begin{align}
 c = 1 + 6 \hat Q ^2 + 2 + 2 \cdot (2N-2M) + 1 +
 \frac{((N-M)^2 - 1) (k-2)}{k -2 - N+M} ~,
\end{align}
which is the same as above.
Note that the central charge for one pair $(p_a,\theta_a)$ is $-2$.

\subsection{Correspondence to a quasi-superconformal theory}

In subsection \ref{BPalgebra} we have studied a relation between
correlators of SL(3) WZNW model and a theory with $W_3^2$ symmetry.
Here we would like to derive similar relations involving the
SL$(2+N|M)$ WZNW model. In order to obtain an explicit formula,
we specify the form of vertex operator, see \eqref{sl3vertexop}.
In the supergroup cases analyzed in \cite{HS2,CHR}, it was useful
to express the fermions in bosonized language.
{}From this experience we again enlarge the Hilbert
space by utilizing the bosonization formula
\begin{align}
 \beta^a_i (z) = - \partial \xi^a_i e^{-  X^a_i} ~, \quad
 \gamma^a_i (z) = \eta^a_i  e^{X^a_i} ~, \quad
 p^a_i (z) = e^{ i Y^a_i} ~, \quad
 \theta^a_i (z) =   e^{ - i Y^a_i} ~,
\end{align}
where the operator products are
\begin{align}
 \xi^a_i (z) \eta^b_j (0) = \delta_{a,b} \delta_{i,j}\frac{1}{z} ~, \qquad X^a_i(z) X^b_j(0) \sim
 Y^a_i(z) Y^b_j(0) \sim - \delta_{a,b} \delta_{i,j} \ln z ~.
\end{align}
Analogous expressions hold for the barred quantities.
Then the vertex operators can be defined as
\begin{align} \label{vertexslnm}
 V^{t^a_i,s^a_{\hat i}}_{j,L} (\mu | z) &=  \mu ^{j + 1 - \frac{1}{2} \sum_{a,i} t^a_i
 + \frac{1}{2} \sum_{a,\hat i} s^a_{\hat i}}
 \bar \mu ^{j + 1 - \frac{1}{2}\sum_{a,i}  \bar t^a_i
+ \frac{1}{2} \sum_{a,\hat i} \bar s^a_{\hat i}} \\ & \cdot
 e^{ t^a_i X^a_i + \bar t^a_i \bar X^a_i + i s^a_{\hat i} Y^a_{\hat i}
+ i \hat s^a_{\hat i} \hat Y^a_{\hat i} }
 e^{\mu \gamma - \bar \mu \bar \gamma}
  e^{2b(j+1) \phi} V^\text{SL$(N|M)$}_L (q) ~, \nonumber
\end{align}
where $L$ labels the representation of sl$(N|M)$.
Other vertex operators may be obtained by applying
$\xi^a_i,\eta^a_i$ and their derivatives.
The correlation functions are computed as in \eqref{correlator}.

In order to obtain relations to a reduced theory, we follow the
strategy adopted in \cite{HS} as in the SL(3) case.
We integrate out $\gamma$, then
the field $\beta$ is replaced by
\begin{align}
 \sum_{\nu=1}^n \frac{\mu_\nu}{z-z_\nu} = u \frac{\prod_{l=1}^{n-2} (z - y_l)}
 {\prod_{\nu=1}^n (z - z_\nu) } = u {\cal B} (y_l , z_\nu ; z) ~,
\end{align}
and similarly for $\bar \beta$. Now the action includes the functions
${\cal B}$ and $\bar {\cal B}$, and the functions can be removed
by the shift of fields $\phi,X^a_i,Y^a_{\hat i}$ as
\begin{align} \label{shiftslnm}
  \phi  + \frac{1}{2b} \ln |u{\cal B} |^2 \to \phi ~, \qquad
  X^a_i  + \frac12 \ln u{\cal B}  \to X^a_i ~, \qquad
  Y^a_{\hat i}  + \frac{i}{2} \ln u{\cal B}  \to Y^a_{\hat i} ~.
\end{align}
After some manipulations, we arrive at the relation among the
correlators as
\begin{align}\label{eq:CorrRela}
\langle \prod_{\nu=1}^n V^{ {t^a_i}_\nu ,{s^a_{\hat i}}_\nu }_{j_\nu,L_\nu} (\mu_\nu | z_\nu)  \rangle  = \delta^{(2)} (\sum_{\nu=1}^n \mu_\nu )
 | \Theta_n |^2  \langle \prod_{\nu=1}^n
V^{ {t^a_i}_\nu -1/2 , {s^a_{\hat i}}_\nu +1/2}_{b(j_\nu + 1) + 1/2b,L_\nu}  (z_\nu)
 \prod_{l=1}^{n-2} V^{ 1/2 , -1/2}_{-1/2b,0} ( y_l)
 \rangle  ~.
\end{align}
The action for the right hand side is
\begin{align} \label{actionlslnm}
 S & = S^\text{WZNW}_{k-2}[q] + \frac{1}{2 \pi} \int d^2 z \Bigl[
 \bar \partial \Phi \partial \Phi +
\bar \partial \phi \partial \phi + \frac{Q}{4} \sqrt{g} {\cal R} \phi
+ \sum_{a=1}^2 ( P_a \bar \partial \Theta_a^t + \bar P_a \partial \bar \Theta_a^t  )
  \nonumber \\
 & + \frac{1}{k} e^{2 b \phi}  - \frac{1}{k} ( P_1 + \zeta \Theta_2 ) q \zeta
 (\bar P_1 - \bar \Theta_2 )^t e^{ b (\phi - \eta \Phi ) }
 - \frac{1}{k} \bar P_2 q ^{-1} \zeta
 \bar P_2^t e^{ b ( \phi + \eta \Phi ) }\Bigr]
\end{align}
with $Q = \hat Q + b^{-1}$.
The vertex operators are
\begin{align}\label{vertexlslnm}
V^{ t^a_i , s^a_{\hat i}}_{\alpha,L} (z) = e^{ t^a_i X^a_i + \bar t^a_i \bar X^a_i+
 i s^a_{\hat i} Y^a_{\hat i} + i \hat s^a_{\hat i} \hat Y^a_{\hat i} }
e^{2 \alpha \phi} V^\text{SL$(N|M)$}_{L} (q) ( z) ~,
\end{align}
and the pre-factor is
\begin{align}
  \Theta_n = u \prod_{i < j}^n (z_i - z_j)^{\frac{1}{2b^2} + \frac{N-M}{2}}
   \prod_{ p < q}^{n-2} (y_p - y_q)^{\frac{1}{2b^2} + \frac{N-M}{2}}
   \prod_{i =1}^n \prod_{p =1}^{n-2} (z_i - y_p)^{ - \frac{1}{2b^2} - \frac{N-M}{2}} ~.
\end{align}
The result for $N=0$ reproduces the one in \cite{CHR}.
As mentioned in subsection \ref{BPalgebra}, there are extra insertions
of operators with the identity representation $L=0$ at $y_l$,
and shifts of parameters of vertex operators at $z_\nu$.

In the case of SL(3), the reduced theory consists of a free theory and
a theory with $W_3^2$ symmetry. We have similar decoupling in this
generalization. If we rotate $P_a,\Theta_a$ as
\begin{align}
\Theta_1 - \zeta P_2 \to \Theta_1 ~, \quad
\Theta_2 + \zeta P_1 \to \Theta_2 ~, \quad
\bar \Theta_1 + \bar P_2 \to \bar \Theta_1 ~, \quad
\bar \Theta_2 - \bar P_1 \to \bar \Theta_2 ~,
\end{align}
then the action becomes
\begin{align}
&  S[q,\Phi,\phi,\psi_a] =
S^\text{WZNW}_{k-2}[q] + \frac{1}{2 \pi} \int d^2 z \Bigl[
\bar \partial \Phi \partial \Phi +
\bar \partial \phi \partial \phi + \frac{Q}{4} \sqrt{g} {\cal R} \phi
 \\
 &  \qquad \qquad + \sum_{a=1}^2 ( P_a \bar \partial \Theta_a^t + \bar P_a \partial \bar \Theta_a^t  )
+ \frac{1}{k} e^{2 b \phi}  + \frac{1}{k} \zeta \Theta_2 q \zeta
  \bar \Theta_2 ^t e^{ b (\phi - \eta \Phi ) }
 - \frac{1}{k} \bar P_2 q ^{-1} \zeta
 P_2^t e^{ b ( \phi + \eta \Phi ) }\Bigr] ~.\nonumber
\end{align}
We thus see that $P_1,\Theta_1$ is a free decoupled system. As in the SL(3) case it is also here an outstanding problem how to rewrite the vertex operators in terms of the rotated variables. For $M=0$, the symmetry of the theory should be the quasi-superconformal
symmetry discussed in \cite{Romans} and for $M \neq 0$ it is a generalization of it.
Utilizing our free field realization, we could
construct the currents for the symmetry as for SL(3) case.
Furthermore, it should be possible to apply the above analysis to more generic cases in page 14 of \cite{KW}.

\section{The product embedding \underline{$2N$} $ =N$\underline{$2$}}
\label{product}

In the previous section, all $n$-point functions of WZNW models on SL$(2+N|M)$ on a sphere
are found to be written in terms of reduced theories with some specific $W$-symmetry.
Since different $W$-algebras can be obtained though the Hamiltonian reduction by using
different sl(2) embeddings, one may wonder whether we can establish relations between
correlators involving more generic $W$-algebras.
In this section, we give an example corresponding to the simple product embedding of
sl(2) as \underline{$2N$} $ =N$\underline{$2$}.
More generic cases are under investigation.

\subsection{SL$(2N)$ WZNW action}

We start from the action of SL$(2N)$ WZNW model.
In order to adopt the reduction procedure in \cite{HS}, we should find out
a proper free field realization of the WZNW model.
Let us consider sl$(2N)$ and the following embedding of sl(2) (in block matrix form)
\begin{align}\label{}
    t^z=\frac{1}{2} \left (
 \begin{array}{c|c}
  \mathbb{I}_N & 0  \\ \hline
   0 & -\mathbb{I}_N
 \end{array} \right)\, ,\qquad t^+= \left (
 \begin{array}{c|c}
  0 & \mathbb{I}_N   \\ \hline
   0 & 0
 \end{array} \right)\, ,\qquad t^-= \left (
 \begin{array}{c|c}
  0 & 0 \\ \hline
   \mathbb{I}_N & 0
 \end{array} \right)\, .
\end{align}
The commutant of this sl(2) is sl$(N)$ generated by
\begin{align}\label{}
    t_{\text{sl}(N)}^\mathbb{A}= \left (
 \begin{array}{c|c}
  \mathbb{A} & 0 \\ \hline
   0 & \mathbb{A}
 \end{array} \right)\, .
\end{align}
The splitting of the fundamental representation of sl$(2N)$ can now be written as
\begin{align}\label{}
    \underline{2N}_{\text{sl}(2N)}=\underline{N}_{\text{sl}(N)}\, \underline{2}_{\text{sl}(2)} \, .
\end{align}

We get a 3-grading of the algebra from the eigenvalues of $t^z$. In this case upper left block is $+1$, lower left $-1$ and the diagonal blocks are zero graded. Following \cite{Bais:1990bs} we make the standard Gauss decomposition obeying this grading
\begin{align}\label{}
    g=\left (
 \begin{array}{c|c}
  \mathbb{I}_N & 0 \\ \hline
   \gamma & \mathbb{I}_N
 \end{array} \right)e^{2\phi t^z}\left (
 \begin{array}{c|c}
  g_1 & 0 \\ \hline
   0 & g_2
 \end{array} \right)\left (
 \begin{array}{c|c}
  \mathbb{I}_N & \bar \gamma \\ \hline
   0 & \mathbb{I}_N
 \end{array} \right)\ ,
\end{align}
where $g_1$ and $g_2$ are SL$(N)$ matrices.
The action then takes the form (with $k$ having opposite sign as in the compact case)
\begin{align}
 S_k^{\textrm{WZNW}} [g] &= \frac{k}{4 \pi} \int_{\Sigma} d ^2z
 \langle g^{-1} \partial g , g^{-1} \bar \partial g \rangle
 + \frac{k}{2 4 \pi} \int_{B} \langle g^{-1} d g
  [g^{-1} d g , g^{-1} d g ] \rangle \nonumber \\
  &= S_k^{\textrm{WZNW}}[g_1]+S_k^{\textrm{WZNW}}[g_2]+\frac{k}{2 \pi} \int d^2 z\, N\del\phi\delbar\phi+ e^{2\phi}\, \tr(\delbar \gamma g_1\del\bar\gamma g_2^{-1})
\end{align}
as is easily found using the Polyakov-Wiegmann identity \eqref{PWid}.
We then go to a first order formalism by introducing the $N\times N$ matrices $\beta,\bar\beta$
\begin{align}\label{2Naction}
    S_k [\phi,g_1,g_2,\gamma,\bar\gamma,\beta,\bar\beta]=& S_{k-N}^{\textrm{WZNW}}[g_1]+S_{k-N}^{\textrm{WZNW}}[g_2]+\frac{1}{2 \pi} \int d^2 z\, \del\phi\delbar\phi-\frac{N^2 b}{4}\sqrt{g}\mathcal{R}\phi\nonumber \\
   &+\frac{1}{2 \pi} \int d^2 z\, \tr(\beta\delbar\gamma+\bar\beta\del\bar\gamma-\frac{1}{k}e^{-2\phi}\bar \beta g^{-1}_1\beta g_2) ~ .
\end{align}
Here we have introduced $b=1/\sqrt{N(k-2N)}$ and rescaled $\phi\mapsto b\phi$.
To find the renormalization we have used (in the old $\phi$)
\begin{align}\label{}
    \beta=k e^{2\phi}g_1\del\bar\gamma g_2^{-1}
\end{align}
and that the contribution from the path integral measure for each of the $N^2$ $\beta,\gamma$ systems then is
\begin{align}
\label{anomaly1}
\delta S= - \frac{1}{\pi} \int d^2 z \partial \phi \bar \partial \phi
  - \frac{1}{8 \pi} \int d^2z  \sqrt{g} {\cal R } \phi ~,
\end{align}
and a similar analysis can be made using $g_1$ and $g_2$ in their Cartan sector. Since they have unit determinant, we do not get any background charge for the SL($N$) factors.
We can check that the central charge is correctly reproduced as
\begin{align}\label{}
    c&=2c_{sl(N)_{k-N}}+N^2 c_{\beta\gamma}+c_{\phi}=2\frac{(N^2-1)(k-N)}{k-N-N}+2N^2+1+6\frac{N^3}{(k-2N)}\nonumber \\
    &=\frac{((2N)^2-1)k}{k-2N} ~.
\end{align}
The currents in this hybrid first order formulation are constructed in \cite{Bais:1990bs} (they introduce this name).
They are of the form
\begin{align}\label{eq:currentsl2n}
    J=\left (\begin{array}{c|c}
  H^1 & \beta \\ \hline
   J^+ & H^2
 \end{array}\right ) ~,
\end{align}
where $H^i$ contain bilinears in $\beta$ and $\gamma$, while $J^+$ needs cubic terms.

\subsection{The reduced action}

We will now see how to obtain the reduced action without making constraints on the momenta of the vertex operators. Due to the form of the currents we can choose a simple $\mu_{ij}$ basis, $i,j=1,\ldots,N$, of the vertex operators
\begin{align}
V_{\alpha,r_1,r_2,\mu_{ij},\bar\mu_{ij}} (z) =e^{\mu_{ij}\gamma^{ji}-\bar\mu_{ij}\bar\gamma^{ji}}
e^{2 \alpha b \phi} V^\text{SL($N$)}_{r_1} (g_1) (z)V^\text{SL($N$)}_{r_2} (g_2) (z) ~,
\end{align}
where $r_1,r_2$ label two sl($N$) representations.
As in the previous cases, the interaction term in the action \eqref{2Naction}
does not depend on $\gamma, \bar \gamma$. Since the kinetic terms in the Lagrangian are of the form $\beta_{ij}\delbar\gamma^{ji}$,
we can thus integrate out $\gamma$ and $\bar\gamma$ in order to obtain
\begin{align}\label{}
    \beta_{ij}\mapsto \sum_{\nu=1}^{n} \frac{\mu^\nu_{ij}}{z-z_\nu} = u_{ij} \frac{\prod_{l=1}^{n-2} (z - y^l_{ij})}
 {\prod_{\nu=1}^n (z - z_\nu) } =  {\cal B}_{ij} (y^l_{ij} , z_\nu ; z)\ ,
\end{align}
where $n$ is the number of vertex operators.
Likewise we have
\begin{align}\label{}
    \bar\beta_{ij}\mapsto - \sum_{\nu=1}^{n} \frac{\bar\mu_{ji}}{\bar z-\bar z_\nu} = - \bar u_{ji} \frac{\prod_{l=1}^{n-2} (\bar z - \bar y^l_{ji})}
 {\prod_{\nu=1}^n (\bar z -\bar z_\nu) } = - \bar{\cal B}_{ji} (\bar y^l_{ij} ,\bar z_\nu ;\bar z)\ .
\end{align}

The interaction term then takes the form
\begin{align}\label{eq:interactionwithmatrices}
 S_{\textrm{int}} =  \frac{1}{2 \pi} \int d^2 z\, \tr(\frac{1}{k}e^{-2b\phi}{\cal B}^\dagger g^{-1}_1{\cal B} g_2) \ .
\end{align}
Here we would like to remove ${\cal B}$ from the action by making use of field redefinitions.
Since $g_1,g_2 \in \text{SL}(N)$, the SL$(N)$ part of ${\cal B}$ may be removed by the
change of $g_1,g_2$, which will be discussed later.
First, the determinant of ${\cal B}$ can be rescaled to one
by a translation in $\phi$. This will give extra insertions of $\phi$ where the determinant is zero or infinite:
\begin{align}\label{}
    \phi&=\varphi+\frac{1}{2b}\ln \left|\frac{d(z)^{1/N}}{\prod_{\nu=1}^n (z - z_\nu)}\right|^2\ ,\nonumber \\
    d(z)&:=\det \{\prod_{l=1}^{n-2} u_{ij}(z - y^l_{ij})\}_{i,j=1,\ldots,N}=\det \{u_{ij}\}\prod_{m=1}^{N(n-2)} (z-d_m)\ ,
\end{align}
assuming $\det \{u_{ij}\}\neq0$. Here we have just used that the determinant is a polynomial in $z$ of degree $N(n-2)$ and we have denoted the zeroes $d_m$.

We will now use that
\begin{align}\label{eq:loglineintegral}
    \int d^2 z\, \del\delbar \ln z \, f(z,\bar z) = -2\pi\int_0^\infty d z\, \del f(z,\bar z)\ , \nonumber \\
    \int d^2 z\, \del\delbar \ln \bar z\,  f(z,\bar z) = -2\pi\int_0^\infty d \bar z\, \delbar f(z,\bar z)\ ,
\end{align}
where the paths are along the branch cut of the logarithm, and $f$ is a function which is smooth in a neighborhood of the branch cut. Adding the two integrals gives a path integral over a complete derivative just giving the difference of the values at the endpoints. The insertions we get from the transformation of $\phi$ are then
\begin{equation}\label{eq:extrau1insertions}
\begin{split}
 &\exp(- \frac{1}{2 \pi} \int d^2 z\, \del\phi\delbar\phi)\propto\\
 &\qquad\qquad\qquad\qquad\exp(- \frac{1}{2 \pi} \int d^2 z\, \del\varphi\delbar\varphi)\prod_{m=1}^{N(n-2
   )}e^{\frac{1}{bN}\varphi(d_m)}\prod_{\nu=1}^{n}e^{-\frac{1}{b}\varphi(z_\nu)}e^{\frac{2}{b}\varphi(\infty)}\ .
\end{split}
\end{equation}
Here we have left out a constant which also contains zeroes.
The zeroes precisely cancel the infinities coming from creating the new exponential vertex operator since they need normal ordering.
The insertions of $\varphi$ at infinity can be seen as an addition of the background charge. We can see this precisely using
\begin{align}\label{}
    \sqrt{g} {\cal R } =-4\del\delbar\ln |\rho|^2 ~.
\end{align}
Since $\rho=1$ everywhere except at infinity, $z'=1/z=0$, we have zero curvature in the whole plane, but a delta function at infinity
\begin{align}\label{}
    \sqrt{g} {\cal R } =-4\del\delbar\ln |z'|^{-2}=16\pi\delta^{(2)}(z') ~.
\end{align}
We thus get the background charge
\begin{align}\label{}
    Q_{\varphi}=-bN^2-1/b\ ,
\end{align}
In appendix \ref{DSreduction} we analyze the Drinfeld-Sokolov reduction and find the correct improved Virasoro tensor and this result compares precisely, see eq. \eqref{eq:Treducedcorrect}. Note there will also be a constant term coming from the background charge due to the constant translation of $\phi$.

One last technicality needs to be considered before continuing.
When we get the extra insertions in eq. \eqref{eq:extrau1insertions} some of them are inserted in the same point as the original vertex operators.
This will create infinities from placing two operators in the same point,
but these are exactly canceled by the zeroes coming from the change of $\phi$ in the existing vertex operator which will also give some extra constant term.
We will see something similar happen later.

The action is now
\begin{align}\label{}
    S_k [\phi,g_1,g_2]=& S_{k-N}^{\textrm{WZNW}}[g_1]+S_{k-N}^{\textrm{WZNW}}[g_2]+\frac{1}{2 \pi} \int d^2 z\, \del\varphi\delbar\varphi+\frac{Q_{\varphi}}{4}\sqrt{g}\mathcal{R}\varphi\nonumber \\
   &+  \frac{1}{2 \pi} \int d^2 z\, \tr(\frac{1}{k}e^{-2b\varphi}{\cal B'}^\dagger g^{-1}_1{\cal B'} g_2)\ ,
\end{align}
where
\begin{align}\label{}
 {\cal B}'_{ij}&=u_{ij} \frac{\prod_{l=1}^{n-2} (z - y^l_{ij})}
 {\big(\det \{u_{i'j'}\}\prod_{m=1}^{N(n-2)} (z - d_m)\big)^{1/N} } ~.
\end{align}
The correlator becomes
\begin{multline}\label{}
 \langle \prod_{\nu=1}^n V_{\alpha^\nu,r^\nu_1,r^\nu_2,\mu^\nu_{ij},\bar\mu^\nu_{ij}} (z_\nu) \rangle_{SL(2N)}=\nonumber \\
|\Theta_\varphi|^2 \langle \prod_{\nu=1}^n |\det \{\mu^\nu_{ij}\}|^{2\alpha^{\nu}/N} e^{(2 \alpha^\nu b-\frac{1}{b}) \varphi} V^\text{SL($N$)}_{r^\nu_1} (g_1)V^\text{SL($N$)}_{r^\nu_2} (g_2) (z_\nu)\prod_{m=1}^{N(n-2
   )}e^{\frac{1}{bN}\varphi(d_m)} \rangle
\end{multline}
where
\begin{align}\label{}
    \Theta_{\varphi}= \det \{u_{ij}\}^N\left(\frac{\prod_{m, m'=1,m\neq m'}^{N(n-2)}(d_m-d_{m'})^{1/N^2}\prod_{\nu, \nu'=1,\nu\neq \nu '}^n(z_\nu-z_{\nu'})}{\prod_{m=1}^{N(n-2)}\prod_{\nu=1}^{n}(d_m-z_\nu)^{1+1/N}} \right)^{\frac{1}{4b^2}}\ .
\end{align}
The first factor is due to the background charge.

To proceed we now need to absorb the determinant one matrices ${{\cal B'}, {\cal B}'}^{\dagger}$ in a redefinition of $g_1,g_2$. This will give extra insertions, and to see this explicitly we consider the case $4=2+2$.

\subsection{The example $N=2$}

To make the transformation of $g_1,g_2$, we suggest to do it in terms of the simple steps given in appendix \ref{non-hol}.
As mentioned there, there are in principle many ways to do combine these steps, and correspondingly we would get different expressions for the same amplitude. We have not yet shown that these different expressions actually are equal, but reserve it for future research. However, it turns out that there is a particular choice where we can write all new operators in the correlators as insertions in points using bosonization, and we can avoid some very complicated insertions.

To do this we parameterize our SL$(2)$ fields as
\begin{align}
    g_1&=\left (
 \begin{array}{cc}
  1 & -\gamma_1 \\
   0 & 1
 \end{array} \right)\left (
 \begin{array}{cc}
  e^{-\phi_1} & 0 \\
   0 & e^{\phi_1}
 \end{array} \right)\left (
 \begin{array}{cc}
  1 & 0 \\
   -\bar \gamma_1& 1
 \end{array} \right)\ ,\nonumber \\
 g_2&=\left (
 \begin{array}{cc}
  1 & 0 \\
   \gamma_2 & 1
 \end{array} \right)\left (
 \begin{array}{cc}
  e^{\phi_2} & 0 \\
   0 & e^{-\phi_2}
 \end{array} \right)\left (
 \begin{array}{cc}
  1 & \bar \gamma_2 \\
   0 & 1
 \end{array} \right)\ ,
\end{align}
i.e. $g_1$ has been parameterized transposed inverse compared to appendix \ref{non-hol}.
We now use \eqref{eq:matrixformulations} to write
\begin{align*}
    {\cal B'}=&\Bigg(\begin{array}{cc}
  1 & \frac{u_{12}\prod_{l=1}^{n-2} (z - y^l_{12})}{u_{22}\prod_{l=1}^{n-2} (z - y^l_{22})} \\
   0 & 1
 \end{array} \Bigg)\Bigg(\begin{array}{cc}
  \frac{\big(\det \{u_{i'j'}\}\prod_{m=1}^{N(n-2)} (z - d_m)\big)^{1/N} }{u_{22}\prod_{l=1}^{n-2} (z - y^l_{22})}
 & 0 \\
   0 & \frac{u_{22}\prod_{l=1}^{n-2} (z - y^l_{22})}
 {\big(\det \{u_{i'j'}\}\prod_{m=1}^{N(n-2)} (z - d_m)\big)^{1/N} }
 \end{array} \Bigg)\\
 &\Bigg(\begin{array}{cc}
  1 & 0 \\
   \frac{u_{21}\prod_{l=1}^{n-2} (z - y^l_{21})}{u_{22}\prod_{l=1}^{n-2} (z - y^l_{22})} & 1
 \end{array} \Bigg) ~.
\end{align*}
Thus we make the following field change
\begin{equation}\label{}
    g_1=A g'_1 A^\dagger\ ,\qquad g_2=B g'_2 B^\dagger
\end{equation}
where
\begin{align*}
 A&=\Bigg(\begin{array}{cc}
  1 & \frac{u_{12}\prod_{l=1}^{n-2} (z - y^l_{12})}{u_{22}\prod_{l=1}^{n-2} (z - y^l_{22})} \\
   0 & 1
 \end{array} \Bigg)\left(\begin{array}{cc}
  \frac{\big(\det \{u_{i'j'}\}\prod_{m=1}^{N(n-2)} (z - d_m)\big)^{1/2N} }{\big(u_{22}\prod_{l=1}^{n-2} (z - y^l_{22})\big)^{1/2}}
 & 0 \\
   0 & \frac{\big(u_{22}\prod_{l=1}^{n-2} (z - y^l_{22})\big)^{1/2}}{\big(\det \{u_{i'j'}\}\prod_{m=1}^{N(n-2)} (z - d_m)\big)^{1/2N} }
 \end{array} \right)\nonumber\\
 &:=\left(\begin{array}{cc}
  1 & \frac{f_{12}}{f_{22}} \\
   0 & 1
 \end{array} \right)\left(\begin{array}{cc}
  \frac{1 }{(f_{22})^{1/2}}
 & 0 \\
   0 & (f_{22})^{1/2}
 \end{array} \right) ~ , \nonumber\\
 B&=\Bigg(\begin{array}{cc}
  1 & 0 \\
   -\frac{f_{21}}{f_{22}} & 1
 \end{array} \Bigg)\left(\begin{array}{cc}
  (f_{22})^{1/2}
 & 0 \\
   0 & \frac{1 }{(f_{22})^{1/2}}
 \end{array} \right) ~.
\end{align*}
Using the notation of appendix \ref{non-hol}, we can now compactly write the total change in the correlators in terms of the currents of $J_i^{\pm,z}$ constructed out of the two sl(2) generators $g_i$:
\begin{equation*}
\begin{split}
 \langle \prod_\nu V(z_\nu) \rangle_{\text{SL}(4)}\ = \ &
 \langle e^{\frac{i}{2 \pi} \oint_{\del R_2} d z\,\left(\ln f_{22}(-J_1^z+J^z_2)\right)+\textrm{conj.}}\\
&\ \
 e^{\frac{i}{2 \pi} \oint_{\del R_1} d z\,\left( \frac{f_{12}}{f_{22}}J_1^+-\frac{f_{21}}{f_{22}}J_2^-+\frac{1}{b}\ln \frac{d(z)^{1/N}}{\prod_{\nu=1}^n (z - z_\nu)}J_{\phi}\right)+\textrm{conj.}}\prod_\nu V(z_\nu)|_{\gamma=\bar\gamma=0} \rangle_{4=2+2},
\end{split}
\end{equation*}
Here $J_\phi=\del\phi$ and $R_1\subset R_2$ are regions containing all the vertex operators, singularities and branch cuts.
The integral around $R_1$ has to be performed first. The OPEs are taken from the $\textrm{SL}(N)\times\textrm{SL}(N)$ theory, and the interaction terms are simply transformed to absorb the $\mathcal{B}$-functions. The evaluation will give the extra insertions and also the field independent factors as explained in appendix \ref{non-hol}.

Firstly the $-\frac{f_{21}}{f_{22}}J_2^-$ transformation simply gives the following insertions in $y_{22}^l$, see eq. \eqref{eq:betainsertion}, and similarly for the transformation $\frac{f_{12}}{f_{22}}J_1^+$ since we have parameterized it dually
\begin{align}\label{eq:betainsertions}
    \prod_{l=1}^{n-2} \exp\left(\frac{u_{21}\prod_{l'=1}^{n-2} (y^{l}_{22} - y^{l'}_{21})}{u_{22}\prod_{l'=1,l'\neq l}^{n-2} (y^l_{22} - y^{l'}_{22})}\beta_2(y^l_{22})+\frac{u_{12}\prod_{l'=1}^{n-2} (y^{l}_{22} - y^{l'}_{12})}{u_{22}\prod_{l'=1,l'\neq l}^{n-2} (y^l_{22} - y^{l'}_{22})}\beta_1(y^l_{22})\right)\ .
\end{align}
Next we perform the transformation $\ln f_{22}(-J_1^z+J^z_2)$. In principle we will get a line integral over the currents, but if we bosonize like in \eqref{eq:bosonization} this will give us extra insertions \eqref{eq:extrainsertionsbosonized}
\begin{align}\label{eq:extrainsertdeltafuncs}
  \prod_{i=1}^2 \Theta_i\prod_{l=1}^{n-2}e^{ (\phi_i/b_i-X_i)(y_{22}^l)}\prod_{m=1}^{N(n-2)}e^{-\frac{1}{N} (\phi_i/b_i-X_i)(d_m)}
\end{align}
where $b_i=(k_i-2)^{-1/2}=(k-N-2)^{-1/2}$ and the  constant factors are
\begin{align}\label{}
    \Theta_i= \frac{\prod_{l,l'=1,l'\neq l}^{n-2} (y_{22}^l - y^{l'}_{22})^{k_i/4}\prod_{m,m'=1,m'\neq m}^{N(n-2)} (d_m - d_{m'})^{k_i/4N^2} }{\prod_{m=1}^{N(n-2)}\prod_{l=1}^{n-2} (d_m - y^{l}_{22})^{k_i/2N}}\ .
\end{align}
Here we have again removed some zeroes which cancel the infinities from creating the new insertions. Further we have to change the $\beta$s in the previous insertions \eqref{eq:betainsertions}. However, this create zeroes which should cancel infinities from when the new insertions \eqref{eq:extrainsertdeltafuncs} are in the same position as the $\beta_i$ insertions. To see this cancellation let us move the new insertions to $y_{22}^l+\epsilon$. We then have (focusing on of the $\beta_2(y^l_{22})$ insertion)
\begin{align}\label{}
     &  \exp\left(\frac{u_{22}\prod_{l''=1}^{n-2} (y^l_{22} - (y^{l''}_{22}+\epsilon))}{\big(\det \{u_{i'j'}\}\prod_{m=1}^{N(n-2)} (y^l_{22} - d_m)\big)^{1/N} }\frac{u_{21}\prod_{l'=1}^{n-2} (y^{l}_{22} - y^{l'}_{21})}{u_{22}\prod_{l'=1,l'\neq l}^{n-2} (y^l_{22} - y^{l'}_{22})}\beta_2(y^l_{22})\right)e^{ (\phi_2/b_2-X_2)(y_{22}^l+\epsilon)}\nonumber \\
 & \qquad    \stackrel{\epsilon\rightarrow0}{\rightarrow}  \         :\exp\left(\frac{u_{21}\prod_{l'=1}^{n-2} (y^{l}_{22} - y^{l'}_{21})}{\big(\det \{u_{i'j'}\}\prod_{m=1}^{N(n-2)} (y^l_{22} - d_m)\big)^{1/N} }\beta_2(y^l_{22})\right)e^{ (\phi_2/b_2-X_2)(y_{22}^l)}:\ ,
\end{align}
since we have the OPEs
\begin{align}\label{}
    (\beta_2)^p(z)e^{-X_2(w)}=(z-w)^{-p}:(\beta_2)^p(z)e^{-X_2(w)}:\ .
\end{align}

We can now write out the relation between WZNW amplitudes and the model with $4=2+2$ $W$-algebra
\begin{multline}\label{}
\langle \prod_{\nu=1}^n V_{\alpha^\nu,r^\nu_1,r^\nu_2,\mu^\nu_{ij},\bar\mu^\nu_{ij}} (z_\nu) \rangle_{\text{SL}(4)}= \\
|\Theta|^2  \langle \prod_{\nu=1}^n |\det \{\mu^\nu_{ij}\}|^{\alpha^{\nu}} e^{(2 \alpha^\nu b-\frac{1}{b}) \varphi} V^\text{SL($2$)}_{r^\nu_1} (A g_1 A^\dagger)V^\text{SL($2$)}_{r^\nu_2} (B g_2 B^\dagger) (z_\nu)\\
\prod_{m=1}^{2(n-2
   )}e^{\frac{1}{2b}\varphi(d_m)}e^{-\frac{1}{2} ((k-4)^{1/2}\phi_1-X_1)(d_m)}e^{-\frac{1}{2} ((k-4)^{1/2}\phi_2-X_2)(d_m)}\\
   \prod_{l=1}^{n-2}:e^{\frac{u_{12}\prod_{l'=1}^{n-2} (y^{l}_{22} - y^{l'}_{12})}{\det \{u_{i'j'}\}^{1/2}\prod_{m=1}^{2(n-2)} (y^l_{22} - d_m)^{1/2} }\beta_1(y^l_{22})}e^{ ((k-4)^{1/2}\phi_1-X_1)(y_{22}^l)}:\\
    :e^{\frac{u_{21}\prod_{l'=1}^{n-2} (y^{l}_{22} - y^{l'}_{21})}{\det \{u_{i'j'}\}^{1/2}\prod_{m=1}^{2(n-2)} (y^l_{22} - d_m)^{1/2} }\beta_2(y^l_{22})}e^{ ((k-4)^{1/2}\phi_2-X_2)(y_{22}^l)}:
   \rangle_{4=2+2} ~,
\end{multline}
where $\Theta$ is
\begin{multline}\label{}
    \Theta=\det \{u_{ij}\}^2 \\
    \frac{\prod_{\nu, \nu'=1,\nu\neq \nu '}^n(z_\nu-z_{\nu'})^{(k-4)/2}\prod_{l,l'=1,l'\neq l}^{n-2} (y_{22}^l - y^{l'}_{22})^{(k-2)/2}\prod_{m,m'=1,m'\neq m}^{2(n-2)} (d_m - d_{m'})^{(k-6)/8} }{\prod_{m=1}^{2(n-2)}\prod_{\nu=1}^{n}(d_m-z_\nu)^{3(k-4)/4}\prod_{m=1}^{2(n-2)}\prod_{l=1}^{n-2} (d_m - y^{l}_{22})^{(k-2)/2}}\ ,
\end{multline}
and the action is
\begin{align}\label{}
    S_k [\phi,g_1,g_2]=& S_{k-2}^{\textrm{WZNW}}[g_1]+S_{k-2}^{\textrm{WZNW}}[g_2]+\frac{1}{2 \pi} \int d^2 z\, \del\varphi\delbar\varphi+\frac{Q_{\varphi}}{4}\sqrt{g}\mathcal{R}\varphi\nonumber \\
   &\pm  \frac{1}{2 \pi} \int d^2 z\, \tr(\frac{1}{k}e^{-2b\varphi} g^{-1}_1 g_2)\ ,
\end{align}
with $b=1/\sqrt{2(k-4)}$ and $Q_{\varphi}=-4b-1/b$.
The currents for the $4=2+2$ $W$-algebra can be written in terms of our free fields explicitly, and the action is invariant under these.

We saw that the procedure already becomes very technical in the simplest example where the $W$-algebra corresponds to the partition $2+2=4$. However, we believe that the analysis generalizes to generic $N$, but leave the analysis for future work.

\section{SL$(2N+M|P)$ WZNW model and $W$-algebras}
\label{extention}

In this section, we would like to sketch a generalization of our results by finding the correspondence between the sl$(2N+M|P)$ WZNW model and the $W$-algebra corresponding to the partition
\begin{equation}
\underline {2N+M|P } \ = \ N\underline{2|0}+M\underline{1|0}+P\underline{0|1}\, .
\label{spartition}
\end{equation}
This will be done on the level of the action without considering the details of the vertex operators.
The Lie superalgebra sl$(2N+M|P)$ is best pictured as supertraceless $(2N+M|P)\times (2N+M|P)$ matrices.
The embedding of sl(2) that provides above decomposition of the fundamental representation is given by
\begin{align}\label{}
    t^z=\frac{1}{2} \left (
 \begin{array}{c|c|c}
  \mathbb{I}_N & 0 & 0 \\ \hline
   0 & -\mathbb{I}_N & 0 \\ \hline
   0 & 0 & 0_{M|P}
 \end{array} \right) , \  t^+= \left (
 \begin{array}{c|c|c}
  0 & \mathbb{I}_N & 0  \\ \hline
   0 & 0 & 0 \\ \hline
   0 & 0 & 0_{M|P}
 \end{array} \right) , \  t^-= \left (
 \begin{array}{c|c|c}
  0 & 0 & 0\\ \hline
   \mathbb{I}_N & 0 & 0 \\ \hline
   0 & 0 & 0_{M|P}
 \end{array} \right) .
\end{align}
The commutant of sl(2) is sl$(N) \oplus$ sl$(M|P) \oplus$ u(1) (except if $M=P+0$, then there is no u(1))
generated by the elements corresponding to the matrix
\begin{align}\label{}
     X\ =\  \left (
 \begin{array}{c|c|c}
  A & 0 & 0 \\ \hline
   0 & A & 0 \\ \hline
   0 & 0 & B
 \end{array} \right) \, .
\end{align}
This means that the $W$-algebra will have sl$(N)$ $\oplus$ sl$(M|P)$ $\oplus$ u(1) current symmetry
extended by some higher-dimensional fields. For example, there will be $2NM$ bosonic dimension-$3/2$ fields and
$2NP$ fermionic ones, and these transform in the tensor product of the fundamental of sl$(N)$ and the anti-fundamental
of sl$(M|P)$ plus its conjugate representation.

The superalgebra decomposes into $t^z$ graded components as follows
\begin{align}\label{}
     Y\ =\  \left (
 \begin{array}{c|c|c}
  A_0 & A_1 & A_{1/2} \\ \hline
  A_{-1} & B_0 & B_{-1/2} \\ \hline
  A_{-1/2} & B_{1/2} & C_0
 \end{array} \right) \, .
\end{align}
Here the components $X_i$ have grade $i$ for $X=A,B,C$.
Let us denote the u(1) generator normalized to have norm one by $t^0$, then the action can be constructed as follows.
Define a group valued field
\begin{align}\label{}
     g\ =\  &\left (
 \begin{array}{c|c|c}
  \mathbb{I}_N & 0 & 0 \\ \hline
   \gamma & \mathbb{I}_N & 0 \\ \hline
   0 & 0 & \mathbb{I}_{M|P}
 \end{array} \right)
\left (
 \begin{array}{c|c|c}
  \mathbb{I}_N & 0 & 0 \\ \hline
  ba/2 & \mathbb{I}_N & b \\ \hline
   a & 0 & \mathbb{I}_{M|P}
 \end{array} \right)
e^{2\phi t^z+Xt^0}\left (
 \begin{array}{c|c|c}
  g_1 & 0 & 0 \\ \hline
   0 & g_2 & 0 \\ \hline
   0 & 0 & g_3
 \end{array} \right)\nonumber   \\
&\times\left (
 \begin{array}{c|c|c}
  \mathbb{I}_N & \bar a\bar b/2 & \bar a \\ \hline
   0 & \mathbb{I}_N & 0 \\ \hline
   0 & \bar b & \mathbb{I}_{M|P}
 \end{array} \right)
 \left (
 \begin{array}{c|c|c}
  \mathbb{I}_N & \bar\gamma & 0 \\ \hline
   0 & \mathbb{I}_N & 0 \\ \hline
   0 & 0 & \mathbb{I}_{M|P}
 \end{array} \right)
\end{align}
then the WZNW action can be rewritten using the Polyakov-Wiegmann identity as
\begin{equation}
\begin{split}
S_k^{\textrm{WZNW}} [g]\ &= \ S_k^{\textrm{WZNW}} [g_1] +S_k^{\textrm{WZNW}} [g_2] + S_k^{\textrm{WZNW}} [g_3]+\\
&\quad +\frac{k}{4\pi}\int d^2z\, (\del X \bar\del X+2N\del\phi\bar\del\phi )+\\
&\quad +\frac{k}{2\pi}\int d^2z\, e^{2\phi} \text{tr}\bigl( (\bar\del\gamma+\frac{1}{2}(\bar\del b a -b\bar\del a))g_1 (\del\bar\gamma+\frac{1}{2}(\bar a\del\bar b-\del\bar a b)g_2^{-1})\bigr)+\\
&\quad + \frac{k}{2\pi}\int d^2z\, e^{\phi+\alpha X} \text{tr}\bigl(g_1 \del\bar ag_3^{-1}\bar\del a  \bigr)+
 \frac{k}{2\pi}\int d^2z\, e^{\phi-\alpha X} \text{tr}\bigl(\bar\del b g_3 \del\bar bg_2^{-1} \bigr)
~.
\end{split}
\end{equation}
Here we defined
\begin{equation}
\alpha \ = \ \sqrt{\frac{2N+M-P}{2N(M-P)}}\, .
\end{equation}
We then pass to a first order formulation by introducing
auxiliary matrix valued fields $\beta,\bar\beta$ and super-matrix fields (partially even partially odd)
$p,\bar p,q,\bar q$. Then after integrating these auxiliary fields in, the action is
equivalent to
\begin{align}
S_k^{\textrm{WZNW}} [g]\ &= \ S_0 + S_{\text{int}} ~, \nonumber \\
S_0 \ &= \ S_{k-N-M+P}^{\textrm{WZNW}} [g_1] +S_{k-N-M+P}^{\textrm{WZNW}} [g_2] + S_{k-2N}^{\textrm{WZNW}} [g_3]+  \nonumber\\
&\quad +\frac{1}{2\pi}\int d^2z\, (\del X \bar\del X+\del\phi\bar\del\phi
+\frac{\hat Q_{\phi}}{4}\sqrt{g}\mathcal{R} \phi
 )+  \\
&\quad + \frac{k}{2\pi }\int d^2z\, \text{tr}(\beta\bar\del\gamma)+\text{tr}(\bar\beta\del\bar\gamma)
+\text{tr}(p\bar\del a)+\text{tr}(\del\bar a\bar p)
+\text{tr}(\bar\del b q)+\text{tr}(\bar q\del\bar b) ~,  \nonumber\\
S_{\text{int}}\ &= \ -\frac{k}{2\pi }\int d^2z\, e^{-2 \delta \phi}\text{tr}(\beta g_2\bar\beta g_1^{-1})+
e^{-\delta \phi-\alpha ' X}\text{tr}((\frac{1}{2}\beta b+p)g_3(\bar p +\frac{1}{2}\bar b\bar\beta)g_1^{-1})+ \nonumber\\
&\quad + e^{-\delta \phi+\alpha ' X}\text{tr}((-\frac{1}{2}\bar\beta \bar a+\bar q)g_3^{-1}(q -\frac{1}{2}a\beta)g_2)  \nonumber
\end{align}
with
\begin{align}
 \delta^{-2} = N (k - 2 N - M + P) ~,  \qquad
\alpha ' = \sqrt{2N}\delta  \alpha ~, \qquad
\hat Q_\phi = \delta N (N + M - P  ) ~.
\end{align}
Quantum corrections come from the Jacobians due to the change of variables as before.
The central charge of the above action is
\begin{align} \nonumber
 c = 1 + 1 + 6 \hat Q_\phi^2 + 2 (N^2 +2 N M - 2 N P) + \frac{2 (N^2-1)(k-N-M+P)}{(k-N-M+P)-N}
 \\
  + \frac{((M-P)^2 - 1) (k-2 N) }{k - 2 N - M + P}
   = \frac{((2N + M - P)^2 - 1) k }{k - 2 N - M + P} ~,
\end{align}
which is that of the SL$(2N+M|P)$ WZNW model as it should be.

Since the interaction terms do not involve $\gamma, \bar \gamma$, we can integrate them out
when the inserted vertex operators are proportional to $\exp (\mu \gamma - \bar \mu \bar \gamma)$.
Then $\beta, \bar \beta$ are replaced by matrices ${\cal B}, - \bar  {\cal B}$, which may be absorbed
by field redefinitions. This field redefinitions will yields extra insertions and shifts of momenta as before.
The reduced action is then
\begin{align}
S \ &= \ S_0 + S_{\text{int}} ~,\nonumber\\
S_0 \ &= \ S_{k-N-M+P}^{\textrm{WZNW}} [g_1] +S_{k-N-M+P}^{\textrm{WZNW}} [g_2] + S_{k-2N}^{\textrm{WZNW}} [g_3]+   \nonumber \\
&\quad +\frac{1}{2\pi}\int d^2z\, (\del X \bar\del X+\del\phi\bar\del\phi
+\frac{ Q_{\phi}}{4}\sqrt{g}\mathcal{R} \phi
 )+ \nonumber \\
&\quad + \frac{k}{2\pi }\int d^2z\,
\text{tr}(p\bar\del a)+\text{tr}(\del\bar a\bar p)
+\text{tr}(\bar\del b q)+\text{tr}(\bar q\del\bar b) ~, \\
S_{\text{int}}\ &= \ -\frac{k}{2\pi }\int d^2z\, e^{-2 \delta \phi}\text{tr}(g_2 g_1^{-1})+
e^{-\delta \phi-\alpha ' X}\text{tr}((\frac{1}{2} b+p)g_3(\bar p +\frac{1}{2}\bar b)g_1^{-1})+ \nonumber\\
&\quad + e^{-\delta \phi+\alpha ' X}\text{tr}((-\frac{1}{2} \bar a+\bar q)g_3^{-1}(q -\frac{1}{2}a)g_2) \nonumber
\end{align}
with
\begin{align}
 Q_\phi = \delta N (N + M - P  ) + \delta^{-1}~.
\end{align}
If we perform proper rotations of  $a,q$ and $b,p$, then half of the system decouples from the rest and becomes free. The remaining part of the action has $W$-algebra symmetry corresponding to the partition \eqref{spartition}.
In this way, correlators of the WZNW model can be written in terms of a theory with
$W$-algebra corresponding to the partition \eqref{spartition}.

\section{Conclusion and discussions}
\label{conclusion}

The present work is a continuation of \cite{HS,HS2,CHR} where path integral methods are used to establish relations
between WZNW theories and their $W$-algebras.
Given a WZNW theory of a Lie algebra, there are $W$-algebras for each inequivalent embedding of sl(2) in the Lie algebra.
Our path integral reduction essentially works if the resulting $W$-algebra is generated by fields of conformal dimension at most two.
There are many $W$-algebras that do not have this property and one of our main future goals is to understand this more general case.
A key example should be the $W_3$-algebra corresponding to the principal embedding of sl(2) in sl(3).
We already found a relation from SL(3) WZNW theory to the Bershadsky-Polyakov algebra, and as a next step we plan to investigate
the relation between the latter and $W_3$ Toda theory. Three different free field realizations of the Bershadsky-Polyakov algebra can be extracted from \cite{Feigin:2004wb} and  one needs to investigate if one of them allows for our path integral techniques. Notice, that to actually construct a correlator correspondence in this case, we also need to solve the problem of the factorization of the vertex operators, expressed in bosonized variables, under the SP(2) rotation \eqref{eq:sp2trans}. There is no guarantee that the transformation of the vertex operators actually is local, and indeed the expression in eq. \eqref{eq:extraghostterms} looks more like a line operator. It is thus an important problem for future research to find the transformed vertex operators, and determine if the correlator dependence holds for point-operators like in the SL(2) case, or a generalization is needed.

Importantly, one can use the correspondence of Liouville theory and the SL(2) WZNW theory to prove the strong-weak duality
between  the two-dimensional Euclidean black hole and sine-Liouville theory \cite{HS3} and its
supersymmetric analogue \cite{Creutzig:2010bt}. The proof proceeds roughly as follows. First, one embeds the gauged WZNW theory in the product
theory SL(2) $\times$ U(1), then one reduces SL(2) to Liouville and finally absorbs the additional degenerate fields in the action.
A possible generalization is to consider gauged WZNW models of type SL$(N)/\mathbb R^{N-1}$. The most important task is to
find a theory that has the same symmetry algebra as the coset theory. In the case of the SL(3) coset, the symmetry algebra has been exhaustively described in Example 7.10 of \cite{Creutzig:2014lsa}, it has 30 generators.

As mentioned in the introduction, the theories with $W_N$-symmetry appear in the context of the AGT correspondence \cite{Alday:2009aq,Wyllard:2009hg}. Interestingly, a relation between the SL$(N)$ WZNW model and  $W_N$ Toda theory can be obtained by inserting surface operators into four-dimensional SU$(N)$ gauge theories \cite{Alday:2010vg,Kozcaz:2010yp}.
In the simplest case with $N=2$, two different interpretations of surface operator lead to the relation between the SL$(2)$ WZNW theory and Liouville theory with extra degenerate insertions \cite{Alday:2010vg}, and a close study was also done in a quite recent paper \cite{Frenkel:2015rda}. The type of surface operator is labeled by the partition of $N$, and then each choice leads to the $W$-algebra labeled by the same partition, such as, the Bershadsky-Polyakov algebra for $N=3$ \cite{Wyllard:2010rp,Wyllard:2010vi}.
The relation to our results definitely deserves further study.

\subsection*{Acknowledgements}

We are grateful to V.~Schomerus for useful discussions.
The work of YH was supported in part by JSPS KAKENHI Grant Number 24740170. 
The work of TC is supported by NSERC grant number RES0019997. The work of PBR was partly funded by AFR grant 3971664 from Fonds National de la Recherche, Luxembourg.

\appendix

\section{Conventions}
\label{conv}

In this appendix the notations for the generators of sl(3) algebra and the sl$(2+N|M)$ superalgebra are summarized.

\subsection{Conventions  for sl(3) algebra}\label{subsec:conv}

The standard notation of sl$(3)$ generators is
\begin{align}
\nonumber
  H_1 &=
 \begin{pmatrix}
  1 & 0 & 0 \\
  0 & -1 & 0 \\
  0 & 0 & 0
 \end{pmatrix} ~,
 &H_2 &=
 \begin{pmatrix}
  0 & 0 & 0 \\
  0 & 1 & 0 \\
  0 & 0 & -1
 \end{pmatrix} ~, \\
  E_1 &=
 \begin{pmatrix}
  0 & 1 & 0 \\
  0 & 0 & 0 \\
  0 & 0 & 0
 \end{pmatrix} ~,
 &E_2 &=
 \begin{pmatrix}
  0 & 0 & 0 \\
  0 & 0 & 1 \\
  0 & 0 & 0
 \end{pmatrix} ~,
 &E_3 &=
 \begin{pmatrix}
  0 & 0 & 1 \\
  0 & 0 & 0 \\
  0 & 0 & 0
 \end{pmatrix} ~, \\
  F_1 &=
 \begin{pmatrix}
  0 & 0 & 0 \\
  1 & 0 & 0 \\
  0 & 0 & 0
 \end{pmatrix} ~,
 &F_2 &=
 \begin{pmatrix}
  0 & 0 & 0 \\
  0 & 0 & 0 \\
  0 & 1 & 0
 \end{pmatrix} ~,
 &F_3 &=
 \begin{pmatrix}
  0 & 0 & 0 \\
  0 & 0 & 0 \\
  1 & 0 & 0
 \end{pmatrix} ~.
 \nonumber
\end{align}
These generators satisfy ($i,j=1,2$)
\begin{align} \label{gmetric}
 [ H_i , H_j ] &= 0 ~, &[ E_i , F_j ] &= \delta_{i,j} H_i ~, \\
 [ H_i , E_j ] &= G_{ij} E_j ~,
 &[ H_i , F_j ] &= - G_{ij} F_j ~,
 &G_{ij} &=
  \begin{pmatrix}
  2 & - 1 \\
  - 1 & 2
\end{pmatrix} ~,\nonumber
\end{align}
and the last commutators are
\begin{align}
\nonumber
 [E_1 , E_2 ] &= E_3 ~,  &[F_1 , F_2 ] &= - F_3 ~,
&[ E_3 , F_3 ] &=  H_1 + H_2 ~, \\
[E_1 , F_3 ] &= -F_2 ~,  &[E_2 , F_3 ] &=  F_1 ~, \\
[F_1 , E_3 ] &= E_2 ~,  &[F_2 , E_3 ] &=  -E_1 ~.
\nonumber
\end{align}

Denote by $\theta$ the longest root, then
\begin{align}
 \theta=\frac{H_1+H_2}{2},\qquad f_{-\theta}=F_3,\qquad e_{\theta}=E_3
\end{align}
forms a SU(2)
\begin{align}
 [e_{\theta},f_{-\theta}]=2\theta,\qquad [\theta,e_{\theta}]=e_{\theta},\qquad [\theta,f_{-\theta}]=-f_{-\theta}\, .
\end{align}
Further we have
\begin{align}
 [\theta,E_1]=\tfrac{1}{2}E_1,\qquad [\theta,E_2]=\tfrac{1}{2} E_2,\qquad [\theta,F_1]=-\tfrac{1}{2}F_1,\qquad [\theta,F_2]=-\tfrac{1}{2}F_2\, .
\end{align}
We thus have a $\mathbb Z_5$-gradation with respect to from the eigenvalues of $\mathrm{ad} \theta$. The zero part contains, besides $\theta$ itself, also
\begin{align}
 \theta^\bot= \frac{H_1-H_2}{2}.
\end{align}
This has non-zero commutators
\begin{align}
  [\theta^\bot,E_1]=\tfrac{3}{2}E_1,\qquad [\theta^\bot ,E_2]=-\tfrac{3}{2} E_2,\qquad [\theta^\bot,F_1]=-\tfrac{3}{2}F_1,\qquad [\theta^\bot,F_2]=\tfrac{3}{2}F_2\, .
\end{align}

\subsection{Free field realization of sl(3) currents}
\label{free}

The sl(3) currents in this first order formalism take the form
\begin{align*}
   &J^{F_1}=\beta_1 - \frac{1}{2} \gamma^2 \beta_3 ~, \qquad
    J^{F_2}=\beta_2 + \frac{1}{2} \gamma^1 \beta_3 ~, \qquad
    J^{F_3}=\beta_3~, \nonumber\\
    &J^{\theta}= \frac{1}{2b}\del\phi - \frac{1}{2} \gamma^1 \beta_1 -
        \frac{1}{2} \gamma^2 \beta_2 - \gamma^3 \beta_3 ~, \qquad
    J^{\theta^\bot}=\frac{3}{2b} \del\phi^\bot  - \frac{3}{2} \gamma^1 \beta_1 +
        \frac{3}{2} \gamma^2 \beta_2 ~, \nonumber\\
    &J^{E_1}=\frac{1}{2b} ( \del\phi +\del\phi^\bot )\gamma^1 -
         \gamma^1 \gamma^1 \beta_1 -
         \frac{1}{4} \gamma^1 \gamma^1 \gamma^2 \beta_3 +
         \frac{1}{2} \gamma^1 \gamma^2 \beta_2 -
         \frac{1}{2} \gamma^1 \gamma^3 \beta_3 -
         \gamma^3 \beta_2 + (k-\frac{1}{2}) \del\gamma^1 ~, \nonumber\\
    &J^{E_2}=\frac{1}{2b}  (\del\phi-3 \del\phi^\bot)\gamma^2  +
         \frac{1}{2} \gamma^1 \gamma^2 \beta_1 +
         \frac{1}{4} \gamma^1 \gamma^2 \gamma^2 \beta_3 -
         \gamma^2 \gamma^2 \beta_2 -
         \frac{1}{2} \gamma^2 \gamma^3 \beta_3 +
         \gamma^3 \beta_1 + (k-\frac{1}{2}) \del\gamma^2 ~, \nonumber\\
    &J^{E_3}=\frac{1}{b} \del\phi \gamma^3 +
         \frac{3}{2b}  \del\phi^\bot \gamma^1 \gamma^2 -
         \frac{1}{2} \gamma^1 \gamma^1 \gamma^2 \beta_1 -
         \frac{1}{4} \gamma^1
           \gamma^1 \gamma^2 \gamma^2 \beta_3 +
         \frac{1}{2} \gamma^1 \gamma^2 \gamma^2 \beta_2 -
         \gamma^1
          \gamma^3 \beta_1 \nonumber\\
          &\qquad - \frac{1}{2}( k-1 ) \gamma^1
           \del\gamma^2 - \gamma^2 \gamma^3 \beta_2 -
         \gamma^3
          \gamma^3 \beta_3 + \frac{1}{2}( k-1) \del\gamma^1
           \gamma^2 + k \del\gamma^3 ~,
\end{align*}
where we have defined $J=k\del g g^{-1}$ due to our opposite sign for $k$, and $J^{t^i}=\tr Jt^i$. Note that the currents are a bit more complex than in the standard decomposition of $g$ corresponding to positive and negative roots. Standard right nested normal ordering is assumed. In particular we have terms with five components. The advantage is a symmetric look along the diagonal of the matrix going from lower left to upper right.

\subsection{Conventions for sl$(2+N|M)$ algebra}
\label{convgen}

We here some of the generators of sl$(2+N|M)$ needed in the text. First we have
\begin{align}
 &Q = \frac{1}{2\sqrt{(N-M)(N-M+2)}}
 \begin{pmatrix}
  ( N-M ) \mathbb{I}_2 & 0 \\
  0 & - 2 \mathbb{I}_{N+M}
 \end{pmatrix} ~, \\
 &E^\pm =
  \begin{pmatrix}
  \sigma^\pm & 0 \\
  0 & 0
 \end{pmatrix} ~, \qquad
  E^0 = \frac{1}{2}
  \begin{pmatrix}
  \sigma^3 & 0 \\
  0 & 0
 \end{pmatrix} ~, \nonumber
\end{align}
where $Q$ is normalized as
$\langle Q ,Q \rangle = \text{str}\, Q Q  = 1/2$.
Other generators are
\begin{align}
 &S^+_{1,i} = e_{i+2,1} ~, \qquad
 S^+_{2,i} = e_{2,i+2} ~, \qquad
 S^-_{1,i} = e_{1,i+2} ~, \qquad
 S^-_{2,i} = e_{i+2 , 2} ~, \\
& F^+_{1,\hat i} = e_{\hat i+2+n,1} ~, \qquad
 F^+_{2,\hat i} = e_{2,\hat i+2+n} ~, \qquad
 F^-_{1,\hat i} = e_{1,\hat i+2+n} ~, \qquad
 F^-_{2,\hat i} = e_{\hat i+2+n , 2} \nonumber
\end{align}
with $(e_{IJ})_{KL} = \delta_{IL} \delta_{JK}$.
Useful commutation relations are
\begin{align}
 &[Q , S^\pm_{1,i}] = \mp \tfrac{1}{2}\eta S^\pm_{1,i} ~, \qquad
 [Q , S^\pm_{2,i}] = \pm \tfrac{1}{2} \eta S^\pm_{2,i}   ~, \qquad
 [E^0 ,  S^\pm_{a,i}] = \pm \tfrac{1}{2} S^\pm_{a,i} ~, \\
 &[Q , F^\pm_{1,\hat i}] = \mp \tfrac{1}{2} \eta F^\pm_{1,\hat i} ~, \qquad
 [Q , F^\pm_{2,\hat i}] = \pm \tfrac{1}{2} \eta F^\pm_{2,\hat i}   ~, \qquad
 [E^0 , F^\pm_{a,\hat i}] = \pm \tfrac{1}{2} F^\pm_{a,\hat i} \nonumber
\end{align}
with $\eta = \sqrt{\frac{N-M+2}{N-M}} $ and $a=1,2$.
With the help of
\begin{align}
  [ S^\pm_{1,i} , S^\pm_{2,j} ] = \mp \delta_{i,j} E^\pm ~, \qquad
  \{ F^\pm_{1,\hat i} , F^\pm_{2,\hat j} \} = \delta_{\hat i,\hat j} E^\pm ~,
\end{align}
we may find
\begin{align}
 & g_{-1/2}^{-1} d g_{-1/2} = \sum_{i=1}^N \left( d \gamma^1_i S^-_{1,i}
  +  d \gamma^2_i S^-_{2,i}
  + \gamma^2_i d \gamma^1_i E^- \right) +
     \sum_{\hat i=1}^M \left( d \theta^1_{\hat i} F^-_{1,\hat i}
  +  d \theta^2_{\hat i} F^-_{2,\hat i}
  - \theta^2_{\hat i} d \theta^1_{\hat i} E^- \right)
~, \nonumber \\
 & d g_{+1/2} g_{+1/2}^{-1} = \sum_{i=1}^N \left( d \bar \gamma^1_i S^+_{1,i}
  +  d \bar \gamma^2_i S^+_{2,i}
  + \bar \gamma^2_i d \bar \gamma^1_i E^+ \right)+
     \sum_{\hat i=1}^M \left( d \bar \theta^1_{\hat i} F^+_{1,\hat i}
  +  d \bar \theta^2_{\hat i} F^+_{2,\hat i}
  + \bar \theta^2_{\hat i} d \bar \theta^1_{\hat i} E^+ \right) ~. \nonumber
\end{align}

\section{Higher genus extension for quasi-superconformal algebra }
\label{genus}

In section \ref{QSCA} we have considered amplitudes on a sphere.
Here we would like to generalize the analysis to those on a compact Riemann
surface $\Sigma$ of genus $g \geq 1$.
We follow the discussions in \cite{HS,HS3}, see also \cite{Fay,Mumford,AMV,VV}.
We denote by $\omega_l$ $(l=1,2,\cdots,g)$ the
holomorphic one forms on $\Sigma$, which are normalized as
\begin{align}
 \oint _{\alpha_k} \omega_l = \delta_{kl} ~, \qquad
 \oint_{\beta_k} \omega_l = \tau_{kl} ~.
\end{align}
Here $(\alpha_k,\beta_k)$ is a canonical basis of homology cycles and $\tau_{kl}$ is
the period matrix of $\Sigma$. In order to express multi-valued functions on $\Sigma$,
we consider the universal cover $\tilde \Sigma$ and
introduce the Abel map $(z_k) = \int^z \omega_k \in \mathbb{C}^g$.
In particular, the Riemann's theta function can be written as
\begin{align} \label{theta}
 \theta_\delta (z | \tau ) =  \sum_{n \in {\mathbb{Z}}^g}
   \exp i \pi [ (n + \delta_1 )^k \tau_{kl} (n + \delta_1 )^l
   + 2 (n+\delta_1 )^k ( z + \delta_2 )_k ]  ~.
\end{align}
Here $\delta_k = (\delta_{1k} , \delta_{2k} )$ with $\delta_{1k},
\delta_{2k} = 0, 1/2$ represents the spin structure along the
homology cycles $\alpha_k$ and $\beta_k$.

We treat the action of $\text{SL}(2+N|M)$ \eqref{actionslnm} in terms of free fields
as the starting point. Then, we can impose the following periodic boundary conditions
for the $\text{SL}(2)$ sub-sector as
\begin{align}
 & \beta ( w_k + \tau_{kl} n^l + m_k | \tau )
   =  e^{2 \pi i n^l \lambda_l} \beta (w_k | \tau ) ~,
  \nonumber \\
 & \gamma ( w_k + \tau_{kl} n^l + m_k | \tau )
   =  e^{ - 2 \pi i n^l \lambda_l} \gamma (w_k | \tau ) ~,
\label{bcg} \\
 & \phi ( w_k + \tau_{kl} n^l + m_k | \tau )
  \ = \ \phi (w_k | \tau ) + \frac{2 \pi n^l {\rm Im} \lambda_l}{b} ~,
\nonumber
\end{align}
and for the free fields as
\begin{align}
 & \beta^a_i ( w_k + \tau_{kl} n^l + m_k | \tau )
   =  e^{ \pi i n^l \lambda_l}\beta^a_i  (w_k | \tau ) ~,
  \nonumber \\
 & \gamma^a_i ( w_k + \tau_{kl} n^l + m_k | \tau )
   =  e^{ -  \pi i n^l \lambda_l}\gamma^a_i (w_k | \tau ) ~,
\\
 & p^a_{\hat i} ( w_k + \tau_{kl} n^l + m_k | \tau )
   =  e^{ \pi i n^l \lambda_l}p^a_{\hat i} (w_k | \tau ) ~,
  \nonumber \\
 & \theta^a_{\hat i} ( w_k + \tau_{kl} n^l + m_k | \tau )
   =  e^{ -  \pi i n^l \lambda_l} \theta^a_{\hat i} (w_k | \tau ) ~.
\nonumber
\end{align}
Similar boundary conditions are imposed for barred expressions.
The twists with $(\lambda_l , \bar \lambda_l)$  are quite specific ones,
but here we consider only the case with these twists.
In the presence of twists, there are no zero mode for $\gamma$ and
$g-1$ zero modes for $\beta$, which are proportional to
$\lambda$-twisted holomorphic differentials $\omega^\lambda_\sigma$
$(\sigma = 1,2,\cdots,g-1)$.
Their explicit expressions may be found in \cite{Bernard}.
Here we fix the coefficients of zero modes
$\varpi_\sigma$ as in \cite{HS},
even though we have to integrate over them for the correlators of WZNW models.
Once the relation is established, we may integrate over the zero modes
depending on the cases considered as in \cite{HS3}.

We study $n$-point functions of vertex operators $V_{j,L}^{t^a_i , s^a_{\hat i}} (\mu | z)$
defined in \eqref{vertexslnm}.
First we integrate $\gamma,\beta$ as before. Then $\beta$ is replaced by the
function as
\begin{equation} \label{bspbg}
\beta (w)  =   \sum_{\nu=1}^n \mu_\nu \sigma_{\lambda} (w ,
z_\nu) + \sum_{\sigma=1}^{g-1} \varpi_\sigma \omega^\lambda_\sigma
(w)  ~ .
\end{equation}
Here $\sigma_\lambda(w,z)$ is a propagator defined in \cite{HS} as
\begin{align}      \label{tp}
\sigma_{\lambda} (w , z)  =
\frac{(h_{\delta}(w))^2}{\theta_{\delta} (\int^w_z \omega | \tau)}
\frac{\theta_\delta (\lambda - \int^w_z \omega| \tau)}
     {\theta_\delta (\lambda | \tau )} ~, \qquad
     ( h_\delta (z)
)^2  = \sum_k \, \partial_k
 \theta_{\delta} (0 |\tau) \omega_k^\lambda (z)
\end{align}
with an odd spin structure $\delta$. Since $\beta (w)$
is a one-differential on a surface of genus $g$, it should
have $2(g-1)$ more zeroes than poles. Thus,
$\beta (w)$ can be expressed as
\begin{align}
 \beta (w)  =  u \frac{\prod_{i=1}^{n+2(g-1)} E(w ,y_i)
 \sigma (w)^2} {\prod_{\nu=1}^n E(w,z_\nu)} ~, \label{condg}
\end{align}
where $E(w,z)$ is the prime form
\begin{align} \label{prime}
 E (z , w )  =   \frac{\theta_\delta ( \int^z_w \omega | \tau)}
                   {h_\delta (z) h_\delta (w)} = u {\cal B}_g (y_i ,z_\nu ; w)~ .
\end{align}
This function has a single zero at $z = w$.
The other function $\sigma (w)$ is a $g/2$-differential,
which has no zeroes and poles. The details can be found
in \cite{HS,VV}. Due to the periodic boundary condition
with $\lambda_l$ \eqref{bcg},
the positions $y_i$ have to satisfy $g$ conditions given in (4.6) of \cite{HS}.

Then the rest would be straightforward. After the shift of fields as in
\eqref{shiftslnm}, the relation becomes
\begin{align}
 \langle \prod_{\nu=1}^n V^{ {t^a_i}_\nu ,{s^a_i}_\nu }_{j_\nu,L_\nu} (\mu_\nu | z_\nu)  \rangle  =
 | \Theta_n^g |^2  \langle \prod_{\nu=1}^n
V^{ {t^a_i}_\nu -1/2 , {s^a_i}_\nu +1/2}_{b(j_\nu + 1) + 1/2b,L_\nu}  (z_\nu)
 \prod_{i=1}^{n+2(g-1)} V^{ 1/2 , -1/2}_{-1/2b,0} ( y_i)
 \rangle  ~,
\end{align}
where the right hand side is computed with the vertex operators \eqref{vertexlslnm}
and the action \eqref{actionlslnm}.  Notice that due to the shift of fields \eqref{shiftslnm},
the fields appearing in the action \eqref{actionlslnm} are single-valued on the
Riemann surface $\Sigma$.
The most complicated calculation is for
$\Theta_n^g$. We deduce it from the known result in \cite{HS,HS3}.
For $\text{SL}(2)$ the pre-factor may be decomposed as
\begin{align}
 |\Theta^g_n|^2 = | \text{det} ' \nabla_\lambda |^{- 2} |\Theta_\text{kin}|^{\frac{1}{b^2} }
 |\Theta_\text{bc}|^2 ~.
\end{align}
The first factor $| \text{det} ' \nabla_\lambda |^{-2}$ comes from the integration
over $\gamma,\beta$ with twist $\lambda$. Here the prime implies that we
do not integrate over zero mode. This factor does not change in our case.
The second factor comes from the kinetic term of $\phi$ and given as
\begin{align}
|\Theta_\text{kin}| &= e^{\frac{3}{4} U_g} \prod_{\nu=1}^n |\sigma (z_\nu)|^{-2}
\prod_{i=1}^{n+2(g-1)} |\sigma (y_i)|^{2}
\\ &\times
 \prod_{\nu < \mu}^n |E(z_\nu , z_\mu)|
\prod_{i < j}^{n+2(g-1)} |E(y_i , y_j)| \prod_{\nu , i}^n |E(z_\nu , y_i)|^{-1}
\nonumber
\end{align}
with
\begin{align}
 U_g = \frac{1}{192 \pi^2} \int d^2 w d^2 y
 ( \sqrt{g} {\cal R} ) (w) ( \sqrt{g} {\cal R} ) (y) \ln |E (w,y)|^2 ~.
\end{align}
Here the power $1/b^2$ comes from $1/2b$ in front of the shift
$1/2b \ln |u {\cal B}|^2$ of $\phi$ in \eqref{shiftslnm}.
{}From \eqref{shiftslnm} we can see that the contributions from one
$X_i^a$ and one $Y_{\hat i}^a$ are $1/(\sqrt{2})^2$ and
$1/(-i\sqrt{2})^2$.%
\footnote{Here we have used a strange notation that operator
product expansions are $\phi (z, \bar z) \phi (0,0)
\sim - 1/2 \ln |z|^2$, $X_i^a (z) X_i^a (0) \sim - \ln z$, and
 $Y_{\hat i}^a (z) Y_{\hat i}^a (0) \sim - \ln z$ as mentioned before.}
Therefore, $1/b^2$ is replaced by $1/b^2 + 2N/2 - 2M /2$
for our case. There are also contributions from the coupling
to the world-sheet curvature and summarized as
\begin{align}
|\Theta_\text{bc}|^2 &= e^{\frac{3}{2} U_g} |u|^{2(1-g)}
\prod_{\nu=1}^n |\sigma (z_\nu)|^{-2}
\prod_{i=1}^{n+2(g-1)} |\sigma (y_i)|^{2} ~ .
\end{align}
The power is related to the product $\hat Q \cdot 1/2b$,
which is $1/2$ for $\text{SL}(2)$.
The background charges of $X_i^a$ and $Y_i^a$ are
$-1/\sqrt2$ and $-i/\sqrt2$, respectively. Therefore,
the factor does not change since
\begin{align}
 \hat Q \cdot \frac{1}{2b} + 2N \cdot \left ( \frac{-1}{\sqrt2} \right )
 \cdot \frac{1}{2 \sqrt2} + 2 M \cdot \left ( \frac{-i}{\sqrt2} \right )
 \cdot \frac{1}{ - 2 i \sqrt2} = \frac{1}{2}
\end{align}
for our case. In total, we have
\begin{align}
 |\Theta^g_n|^2 &= | \text{det} ' \nabla_\lambda |^{-2}
 |\Theta_\text{kin}|^{\frac{1}{b^2} + N - M }
 |\Theta_\text{bc}|^2  \nonumber \\
 &= | \text{det} ' \nabla_\lambda |^{-2}
 |\Theta_\text{kin}|^{k-2}
 |\Theta_\text{bc}|^2  \\
 &=  e^{\frac{3}{4} k U_g}  | \text{det} ' \nabla_\lambda |^{-2} |u|^{2(1-g)}
 \prod_{\nu=1}^n |\sigma (z_\nu)|^{-2k+2}
\prod_{i=1}^{n+2(g-1)} |\sigma (y_i)|^{2k-2} \nonumber
\\ &\times
 \prod_{\nu < \mu}^n |E(z_\nu , z_\mu)|^{k-2}
\prod_{i < j}^{n+2(g-1)} |E(y_i , y_j)|^{k-2} \prod_{\nu , i}^n |E(z_\nu , y_i)|^{2-k} ~.  \nonumber 
\end{align}
Interestingly, the end result does not seem to depend on $N,M$.
If we properly normalize the correlation function by its partition function,
then the pre-factor $|\Theta_n^g|^2$ should be written in a more simple way
as in \cite{HS}.

\section{DS reduction for the embedding \underline{$2N$} $ =N$\underline{$2$}}
\label{DSreduction}

Let us consider the current in eq. \eqref{eq:currentsl2n}. Hamiltonian reduction basically sets all currents to zero that are not highest weight (or lowest in our formulation here) in the SL(2) decomposition, except the field related to $t^+$ which is set to one. Here this means $H_1=H_2$ and $\beta=\mathbb{I}_N$. I.e.
\begin{align}\label{eq:Jcontrained}
    J_{\textrm{fix}}= \left (\begin{array}{c|c}
  J_{N} & \mathbb{I}_N \\ \hline
   T & J_{N}
 \end{array}\right)\ .
\end{align}
$J_N$ and $T$ will be the generators of the symmetries of the reduced model. This means setting $\beta=\mathbb{I}_N$ and solving for $\gamma$ such that $J$ takes the wanted form.
Inserting into the action gives a new action whose interaction term is (this was already done in \cite{Tjin:1991wm})
\begin{align}\label{}
    S_{\textrm{int}}=-\frac{1}{2 k\pi} \int d^2 z\, \tr(e^{-2b\phi} g^{-1}_1 g_2) \, .
\end{align}
Note that this term obviously is symmetric under the SL$(N)$ group that commuted with the embedded SL(2) since $g_1$ and $g_2$ transform in the same way under this.

However, there is a certain gauge freedom allowed. In our language, when we have integrated out $\gamma$ to fix $\beta$ to a matrix of functions, to which value do we then put $\gamma$ in $J$? If we use the value they solve for, we would also get \eqref{eq:Jcontrained}. However if we simply set $\gamma$ to zero we would get (all terms in $J^+$ depend on $\gamma$)
\begin{align}\label{eq:Jcontrained2}
    J_{\textrm{fix}}= \left (\begin{array}{c|c}
  J_1 & \mathbb{I}_N \\ \hline
   0 & J_2
 \end{array}\right)\ .
\end{align}
This is a gauge equivalent constraint also mentioned in \cite{Bais:1990bs}. Here we directly see the two SL($N$) currents, except that they also contain the current for $\phi$. The relation to $J_N$ and $T$ above is
\begin{align}\label{}
    J_N=\frac{1}{2}(J_1+J_2),\qquad T=2\tilde{J}^2 + \del \tilde J,
\end{align}
where $\tilde J=(J_1-J_2)/2-qj\mathbb{I}_N$, and we have explicitly pulled out the U(1) current $j=i\sqrt{2}\del\phi$ with $j(z)j(w)\sim(z-w)^2$, and $q^2=(2N-k)/2N$. The stress-energy tensor of the reduced model is supposed be (remember that we have opposite sign on level)
\begin{align}\label{eq:Timprovedreduced}
    T^{\textrm{Improved}}_{\textrm{red}}&=-\frac{1}{2(k-2N)}\tr J_{\textrm{fix}}^2+\tr{\del J_{\mathrm{fix}}t^z}=-\frac{1}{2(k-2N)}\tr T-\frac{1}{k-2N}\tr J^2_N+2Nq\del j\nonumber \\
    &=-\frac{1}{2(k-2N)}\tr (J_1^2+J_2^2)-\frac{N}{k-2N}q^2j^2+\frac{qN}{2(k-2N)}\del j +2N q\del j\nonumber \\
    &=T^{\text{SL}(N)}_{k-N}(g_1)+T^{\text{SL}(N)}_{k-N}(g_2)-\del\phi\del\phi-(\frac{Nb}{2}+\frac{2}{b})\del^2\phi\ .
\end{align}
There is, however, a factor $2N$ missing in the background charge for the term proportional to $b$. For future reference we now show how to correctly obtain the reduced stress-energy tensor. The improvement term will correspond to the extra correction of the U(1) background charge when we remove the $\beta$ fields from our action. Since we have $k\mapsto-k$ we can write
\begin{align}\label{}
    T^{\textrm{Improved}}_{\textrm{red}}&=-\frac{1}{2(k-2N)}\tr J_{\textrm{fix}}^2-\tr{\del J_{\mathrm{fix}}t^z}+\textrm{background charges},
\end{align}
where we think of $J_{\textrm{fix}}$ in the form \eqref{eq:Jcontrained2} since it is most natural to set the $\gamma$ fields to zero after they have been integrated out. Remembering the shifts in levels, we here have
\begin{align}
    J_{\textrm{fix}}= \left (\begin{array}{c|c}
  J^{g_1}_{k-N}+\frac{1}{Nb}\del\phi\mathbb{I}_N & \mathbb{I}_N \\ \hline
   0 & J^{g_2}_{k-N}-\frac{1}{Nb}\del\phi\mathbb{I}_N
 \end{array}\right)\ .
\end{align}
The extra background charges come for the following reason. When we start with the SL(2$N$) WZNW model we also have terms in the stress energy tensor of the form $\tr(J^+J^-+J^-J^+)/(2(k-2N))$, where plus and minus refer to the upper right and lower left block respectively. $J^-$ are all going to be zero when setting $\gamma=0$, however a normal ordering is required and let us use right nested normal ordering. When setting $\gamma=0$ and hence $J^-=0$, then we want $J^-$ to stand to the right since $J^-$ is then actually evaluated in the point of the normal ordered product and not along some integral. Thus we use
\begin{align}\label{}
    \frac{1}{2(k-2N)}\tr_N J^-J^+&=\frac{1}{2(k-2N)}\tr_N J^+J^-+\frac{1}{2(k-2N)}\tr[J^-,J^+]\nonumber \\
    &=\frac{1}{2(k-2N)}\tr_N J^+J^- + \frac{-2N}{2(k-2N)}\tr_{2N}\del J t^z ~.
\end{align}
Inserting we now get
\begin{align}\label{eq:Treducedcorrect}
    T^{\textrm{Improved}}_{\textrm{red}}&=-\frac{1}{2(k-2N)}\tr J_{\textrm{fix}}^2-\tr{\del J_{\mathrm{fix}}t^z}-\frac{2N}{2(N-2k)}\tr{\del J_{\mathrm{fix}}t^z} \nonumber \\
    &=T^{SL(N)}_{k-N}(g_1)+T^{SL(N)}_{k-N}(g_2)-\del\phi\del\phi-(N^2b+\frac{1}{b})\del^2\phi\ .
\end{align}
This indeed has the correct background charge for the term proportional to $b$, and the term proportional to $1/b$ is exactly produced by removing the $\beta$s.

Decomposing the adjoint representation into sl(2) representations also tells us how the original sl$(2N)$ currents splits into symmetries of the new action. This is done in \cite{Bais:1990bs} and gives
\begin{align}\label{}
    \underline{ad}_{\text{sl}(2N)}\mapsto (\underline{ad}_{\text{sl}(N)}\otimes\underline{1}_{\text{sl}(2)})\oplus(\underline{ad}_{\text{sl}(N)}\otimes\underline{3}_{\text{sl}(2)})\oplus(\underline{1}_{\text{sl}(N)}\otimes\underline{3}_{\text{sl}(2)})\ .
\end{align}
The first term simply gives the sl($N$) currents, the second term gives $N^2-1$ spin-2 fields which transform in the adjoint of sl($N$), and the last terms gives a single spin-2 field which commutes with sl($N$). The spins follow from their eigenvalues under $t^z$ due to the extra background charge that will arise in our formulation when we integrate out $\beta$, or correspondingly due to the choice of an ``improved'' Virasoro tensor. In terms of the current \eqref{eq:currentsl2n} the traceless part of $T$ gives the $N^2-1$ fields and the trace of $T$ the single spin 2 field. In all, we simply have a $W_2$ algebra minimally coupled to sl($N$). The commutators are given in the appendix of \cite{Bais:1990bs} and section 8.2 in \cite{deBoer:1992sy}.

\section{Non-holomorphic field changes in first order SL(2) WZNW model}
\label{non-hol}

Consider the SL(2) WZNW model in first order formalism, i.e. we use parametrization
\begin{align}\label{eq:gmatrix}
    g=\left (
 \begin{array}{cc}
  1 & 0 \\
   \gamma & 1
 \end{array} \right)\left (
 \begin{array}{cc}
  e^\phi & 0 \\
   0 & e^{-\phi}
 \end{array} \right)\left (
 \begin{array}{cc}
  1 & \bar \gamma \\
   0 & 1
 \end{array} \right)\ =\left(\begin{array}{cc}
  e^\phi & \bar \gamma e^\phi \\
   \gamma e^\phi & e^{-\phi}+\gamma\bar\gamma e^{\phi}
 \end{array} \right)\mapsto e^{b\phi}\left(\begin{array}{cc}
  1 & \bar \gamma \\
   \gamma & \gamma\bar\gamma
 \end{array} \right),
\end{align}
with action
\begin{align}\label{}
    S_k [\phi,\gamma,\bar\gamma,\beta,\bar\beta]=& \frac{1}{2 \pi} \int d^2 z\,\del\phi\delbar\phi-\frac{b}{4}\sqrt{g}\mathcal{R}\phi+ \beta\delbar\gamma+\bar\beta\del\bar\gamma-\frac{1}{k}e^{-2b\phi} \beta\bar \beta \ .
\end{align}
Here $b=(k-2)^{-1/2}$. We will only consider the spherical topology in this section.
In the last expression for $g$, we are in the first order formalism where the term $e^{-b\phi}$ is of wrong dimension (it will be a contact term), and has been removed. The currents are ($t^z=\textrm{diag}\{1/2,-1/2\}$)
\begin{align}\label{}
    &J^{z}_B=\frac{1}{b}\del\phi-\normord{\gamma\beta} ~, \qquad
    J^{-}_B=\beta ~,\nonumber\\
    &J^{+}_B=2\frac{1}{b}\del\phi\gamma+k\del\gamma-\normord{\beta\gamma\gamma} ~,
\end{align}
with OPEs
\begin{align}\label{}
    \gamma(z)\beta(w)\sim\frac{1}{z-w}\ , \qquad \phi\phi\sim-\frac{1}{2}\ln|z-w|^2.
\end{align}
Remember here $J=+k \del g g^{-1}$ due to the opposite sign on $k$.

We need to know what happens with the action when make a field redefinition of the form
\begin{align}\label{eq:transgsl2}
    g=A(z)g',
\end{align}
where $A$ is a holomorphic matrix with unit determinant except for poles and branch cuts. The interaction term is invariant under these transformations up to total derivative terms, but there will be effects from the kinetic terms and the measure of integration over the fields.
We split the problem into the three cases corresponding to the grading of the algebra.

\subsubsection*{Changes in $t^-$ direction}

We first consider the transformation
\begin{align}\label{eq:transcase1}
    g=\left (
 \begin{array}{cc}
  1 & 0 \\
   f(z) & 1
 \end{array} \right)g' ~.
\end{align}
In terms of fields this is simply
\begin{align}\label{eq:transgamma}
    \gamma(z)=\gamma'(z)+f(z) ~.
\end{align}
Since this is a translation it will not change the measure and the change simply comes from the kinetic term
\begin{align}\label{}
    S[\gamma]=S[\gamma']+\frac{1}{2 \pi} \int d^2 z\,\beta'\delbar f(z) ~.
\end{align}
We thus only get insertions in poles and along branch cuts. Let us consider
\begin{align}\label{eq:fsimple}
    f(z)=\frac{z-Q}{z-P} ~.
\end{align}
Then we can limit our integration to a small neighborhood of $P$, denoted $R$.
We will here think of $\beta$ as a smooth function that only has poles if we meet insertions. Thus $R$ can be chosen as any area
not containing insertions (such that the partial integration can be made without considering the term with $\delbar\beta$), and we can write (employing Polchinski notation)
\begin{align}\label{}
    S[\gamma]=S[\gamma']+\frac{-i}{2 \pi} \oint_{\del R} d z\,\beta f(z)=S[\gamma']+\frac{-i}{2 \pi} \oint_{\del R} d z\, f(z)J^-,
\end{align}
as is expected from the Polyakov-Wiegmann identity \eqref{PWid}. With \eqref{eq:fsimple} will thus get an insertion of $\beta'$ in $P$:
\begin{align}\label{eq:betainsertion}
    e^{-S[\gamma]}=e^{-S[\gamma']}e^{-(P-Q)\beta'(P)}\ .
\end{align}
The OPE of $\gamma$ with this insertion exactly reproduces the infinity in \eqref{eq:transgamma}
Note that if we consider some correlator $ \langle \prod_i V(z_i) \rangle_{\text{SL}(2)}$ we can write
\begin{align}\label{eq:trans1onvertex}
    \langle \prod_i V(z_i) \rangle_{\text{SL}(2)}= \langle e^{\frac{i}{2 \pi} \oint_{\del R} d z\,f(z)J^-}\prod_i V(z_i) \rangle_{\text{SL}(2)},
\end{align}
where $R$ is now a region encircling all points of insertion (which can include infinity) and poles of $f$. We can deform the contour as long as we do not cross insertions or poles in $f(z)$ due to the Ward identity $\delbar J^-=0$ outside insertions. Since we can close the contour around a point without insertions, the exponential can be evaluated to one giving the identity. The contour integral around an insertion simply generates the transformation $\gamma\mapsto\gamma'$. Further,  remember here that currents generate transformations of $g$ with opposite sign
\begin{align}\label{}
    J^{a}(z)g(w)\sim\frac{-t^a g(w)}{z-w}\ .
\end{align}

\subsubsection*{Changes in $t^z$ direction}

We now consider the transformation
\begin{align}\label{eq:changecase1}
    g(z)=\left (
 \begin{array}{cc}
  f(z) & 0 \\
  0 & 1/f(z)
 \end{array} \right)g'(z) ~.
\end{align}
In terms of fields, the transformation is
\begin{align}\label{eq:fieldchangecase2}
    \phi(z)=\phi'(z)+\frac{1}{b}\ln f(z)\ , \qquad
    \gamma(z)=\frac{1}{f(z)^2}\gamma'(z)\ , \qquad \beta(z)=f(z)^2\beta'(z)\ .
\end{align}
The change in the action is thus (after partial integration over the whole sphere)
\begin{align}\label{}
    S[\phi,\gamma,\beta]=S[\phi',\gamma',\beta']+\frac{1}{2 \pi} \int d^2 z\,\big(\frac{2}{b}\delbar\ln f\del\phi'+f^2\delbar\frac{1}{f^2}\beta'\gamma'  \big)+\textrm{const.}\ ,
\end{align}
where we have ignored a constant that also contains infinities. Let us consider functions of the form
\begin{align}\label{}
    f(z)=\left(\frac{z-Q}{z-P}\right)^{p}\ .
\end{align}
For $2p\in \mathbb{Z}$ we see that the extra terms in $\beta\gamma$ coming from the action are actually zero -- they will have the form of a delta function multiplied with a zero. However, we must also get a contribution from the measure since the change \eqref{eq:fieldchangecase2} tells us that $\beta'$ is forced to have a zero of order $2p$ in $P$ (remember $\beta$ is thought of as a well-behaved function), and further it \emph{can} have a pole in $Q$ (but in principle this can be multiplied by zeroes in $\beta$). Likewise $\gamma'$ is forced to have a zero of order $2p$ in $Q$, and can have a pole in $P$. For $2p$ non-integer we also get zero, but here care has to be taken due to the branch cut. We propose that including effects from the measure we simply get the logarithm version
\begin{align}\label{}
    D\phi D\gamma D\beta e^{S[\phi,\gamma,\beta]}\propto D\phi' D\gamma' D\beta' e^{S[\phi',\gamma',\beta']}e^{\frac{1}{2 \pi} \int d^2 z\big(\frac{2}{b}\delbar\ln f\del\phi-2\delbar\ln f\beta\gamma  \big)}\,.
\end{align}
The integral can only be non-zero in a small region around the branch cut going from $P$ to $Q$ so as before we get
\begin{align}\label{}
    D\phi D\gamma D\beta e^{-S[\phi,\gamma,\beta]}&\propto D\phi' D\gamma' D\beta' e^{-S[\phi',\gamma',\beta']}e^{\frac{i}{2 \pi} \oint_{\delta R} d z\, 2\ln f J^z}\nonumber \\
    &\propto D\phi' D\gamma' D\beta' e^{-S[\phi',\gamma',\beta']}e^{- \int_Q^P  d z\, 2p J^z}\ ,
\end{align}
where the factor two is due to the one half in the definition of $t^z$ and in the last line the integral runs along the branch cut from $Q$ to $P$. As before for correlators we get
\begin{align}\label{eq:trans2onvertex}
    \langle \prod_i V(z_i) \rangle_{\text{SL}(2)}= \langle e^{\frac{i}{2 \pi} \oint_{\del R} d z\,2\ln f(z)J^z}\prod_i V(z_i) \rangle_{\text{SL}(2)} ~,
\end{align}
where $R$ now contains all insertions and poles and branch cuts of $f$. In this operator formalism we can also find the constant of proportionality needed above. We have (away from other insertions)
\begin{align}\label{}
\langle e^{\frac{i}{2 \pi} \oint_{\del R} d z\,2\ln f(z)J^z} \rangle_{\text{SL}(2)}&= \langle e^{\frac{i}{2 \pi} \oint_{\del R'} d z\,2\ln f(z)J^z- \int_Q^P  d z\, 2p J^z}\rangle _{\text{SL}(2)}
\end{align}
where the integral from $Q$ to $P$ is along a curve north of the branch cut and the region $R'$ is containing this curve, but not the branch cut (except at the end points of the curve). The integral over $\del R'$ gives the change $J^z \mapsto J^z+k\del \ln f$ in the integral from $Q$ to $P$, which can also be found using the OPE $J^z J^z\sim \tfrac{-k/2}{(z-w)^2}$. However, it is zero when evaluated away from this. We can thus use the Baker-Campbell-Hausdorff formula in the form $\exp(A+B)=\exp(B)\exp(A)\exp([A,B]/2)$ to get
\begin{align}\label{eq:transcase2withconstants}
\langle e^{\frac{i}{2 \pi} \oint_{\del R} d z\,2\ln f(z)J^z} \rangle_{\text{SL}(2)}= \left(\frac{f(Q)}{f(P)}\right)^{kp} \langle e^{- \int_Q^P  d z\, \left(2p J^z\right)} \rangle_{\text{SL}(2)}\ .
\end{align}
We have also used that the function $\ln f(z)=p\ln (z-Q)/(z-P)$ is holomorphic on the path from $P$ to $Q$ except at the endpoints, such that we have a line integral over a total derivative. The constant contains zeroes which cancels the infinities needed in the normal ordering of the integral over $J^z$.

To see that this is indeed correct, we bosonize the $\beta\gamma$ system
\begin{align}\label{eq:bosonization}
    \beta=e^{-X+Y}\del Y\ ,\qquad \gamma=e^{X-Y},
\end{align}
with kinetic terms
\begin{align}\label{}
    S^{\textrm{gh}}_{\textrm{kin}}=\frac{1}{2 \pi} \int d^2 z \big(\tfrac{1}{2}\del X\delbar X-\tfrac{1}{2}\del Y\delbar Y-\tfrac{1}{8}\sqrt{g}\mathcal{R}(X-Y)\big)\ .
\end{align}
The field transformation \eqref{eq:fieldchangecase2} is now translational without changes in the measure
\begin{align}\label{}
    X(z)=X'(z)-2\ln f(z)\ .
\end{align}
Using \eqref{eq:loglineintegral} we get
\begin{align}\label{eq:extrainsertionscase2}
    D\phi DX DY e^{-S[\phi,X,Y]}=D\phi' DX' DY' e^{-S[\phi',X',Y']}e^{- \int_Q^P  d z\, \left(pk \del \ln f\right)}e^{- \int_Q^P  d z\, \left(2p \del(\phi'/b-X')\right)}\ .
\end{align}
Using $\beta\gamma=\del X$ this gives the same result as before including the constant factor. If we split $X=X_L+X_R$, then as long as we do not have other insertions along the path from $Q$ to $P$ we get a total derivative and the result
\begin{align}\label{eq:extrainsertionsbosonized}
    D\phi DX DY e^{-S[\phi,X,Y]}=D\phi' DX' DY' e^{-S[\phi',X',Y']}\left(\frac{f(Q)}{f(P)}\right)^{kp}e^{2p (\phi'/b-X')(Q)}e^{-2p (\phi'/b-X')(P)}\ ,
\end{align}
where we directly see the need for the zeroes in the constant factor. Note that there is no change from the background charges even for a general function with $f(\infty)\neq1$. Further this has a nice interpretation. Consider $p=1/2$, then
\begin{align}\label{}
    \gamma(z) e^{-X}(w)=\mathcal{O}(z-w)
\end{align}
and thus it is customary to label $e^{-X}(z)=\delta(\gamma(z))$ and likewise $e^{X}(z)=\delta(\beta(z))$. Thus the above extra insertions precisely forces the extra zeroes in $\beta$ and $\gamma$ than we mentioned above.

\subsubsection*{Changes in $t^+$ direction}

Let us finally consider
\begin{align}\label{}
    g=\left (
 \begin{array}{cc}
  1 & f(z) \\
   0 & 1
 \end{array} \right)g'\ ,
\end{align}
where we again think of $f$ in the form \eqref{eq:fsimple}. The strategy to perform this is to take $g\mapsto (g^{-1})^t$, do the transformation \eqref{eq:transcase1} with opposite sign on $f$, and transform back again $g\mapsto (g^{-1})^t$. The transformation to the inverse transposed takes the following form
\begin{align}\label{}
    \phi&\mapsto\phi+\frac{1}{b}\ln (\gamma\bar\gamma)\ ,\nonumber\\
    \gamma&\mapsto-\frac{1}{\gamma}\ ,\nonumber\\
    \beta&\mapsto \normord{\beta\gamma\gamma}-2\frac{1}{b}\del\phi\gamma-k\del\gamma\ -\frac{1}{b^2}\del\bar\gamma\frac{\gamma}{\bar\gamma}\ \simeq -J^+.
\end{align}
The first two equations simply give the wanted transformation if we use the last form of $g$ in \eqref{eq:gmatrix} i.e. removing the term of irrelevant dimension in the first order formalism. The transformation for $\beta$ then follows by keeping the kinetic terms invariant. That $\beta=J^-$ maps to $-J^+$ up to a term that is zero using the equations of motion, simply follows from $J\mapsto-J^t$. Note that there are normal ordering issues when inserting into the kinetic terms. The transformation can however also be done in the bosonized system \eqref{eq:bosonization} where it takes the form
\begin{align}\label{}
    \phi&\mapsto\phi+\frac{1}{b}(X-Y)\ ,\nonumber\\
    X&\mapsto -X+(k-2)(X-Y)+\frac{2}{b}\phi\ ,\qquad Y\mapsto-Y+(k-2)(X-Y)+\frac{2}{b}\phi
\end{align}
followed by flipping the signs $\beta\mapsto-\beta$, $\gamma\mapsto-\gamma$. This gives the same result using
\begin{align}\label{}
    ((\gamma\beta)\gamma)=(\beta(\gamma\gamma))-\del \gamma\ .
\end{align}
Note that in the bosonized version we also see that the background charges are left invariant. This is hard to see in the non-bosonized version since we do not know the changes from the measure.
The total transformation for $\phi$ and $\gamma$ is thus
\begin{align}\label{}
    \phi=\phi'+\frac{1}{b}\ln(1+f \gamma')\ ,\qquad \gamma=\frac{\gamma'}{1+f\gamma'}\ .
\end{align}

For the action we get an extra insertion $\exp((P-Q)\beta(P))$ like in \eqref{eq:betainsertion} after the translation with $-f=-(z-Q)/(z-P)$. However, after the last transformation $g\mapsto(g^{-1})^t$ we thus get
\begin{align}\label{}
    e^{-S[\phi,\gamma,\beta]}=e^{-S[\phi',\gamma',\beta']}e^{-(P-Q)J^+(P)}\ .
\end{align}
In total we can write
\begin{align}\label{eq:trans1onvertex}
    \langle \prod_i V(z_i) \rangle_{\text{SL}(2)}= \langle e^{\frac{i}{2 \pi} \oint_{\del R} d z\,f(z)J^+}\prod_i V(z_i) \rangle_{\text{SL}(2)} ~ ,
\end{align}
where the region $R$ includes infinity.

\subsubsection*{Remarks}

Note that in all the three cases considered, we can write
\begin{align}\label{}
    \langle \prod_i V(z_i) \rangle_{\text{SL}(2)}= \langle e^{\frac{i}{2 \pi} \oint_{\del R} d z\,\tr\left(\ln A(z)J(z)\right)}\prod_i V(z_i) \rangle_{\text{SL}(2)} ~ ,
\end{align}
for the transformation \eqref{eq:transgsl2} since e.g.
\begin{align}\label{}
    \ln \left(\begin{array}{cc}
  1 & f(z) \\
   0 & 1
 \end{array} \right) = \left(\begin{array}{cc}
  0 & f(z) \\
   0 & 0
 \end{array} \right) \ .
\end{align}
The reason is simply that in these cases $\del A$ commutes with $\ln A$.

Also note that even though it is possible to bosonize to get insertions in points in the basic cases, this is not always the case. Consider e.g. first doing the transformation \eqref{eq:changecase1} and then \eqref{eq:transcase1}. Then from the first change we have a line integral over $\beta\gamma$, but the second change will give a translation in $\gamma$ and thus produce an extra line integral over $\beta$ times a function, which cannot directly be bosonized to a derivative of a field. If we think about the first line integral as being delta functions of $\gamma$ and $\beta$ in the endpoints, then the translation will simply give a translation in the argument of the delta functions.

Note that there are many different ways to perform the same total transformation in steps of the three basic operations, e.g.
\begin{align}\label{eq:matrixformulations}
    \left(\begin{array}{cc}
  A & B \\
   C & D
 \end{array} \right) &=\left(\begin{array}{cc}
  1 & 0 \\
   C/A & 1
 \end{array} \right)\left(\begin{array}{cc}
  A & 0 \\
   0 & 1/A
 \end{array} \right)\left(\begin{array}{cc}
  1 & B/A \\
   0 & 1
 \end{array} \right)  ~, \nonumber\\
    \left(\begin{array}{cc}
  A & B \\
   C & D
 \end{array} \right) &=\left(\begin{array}{cc}
  1 & B/D \\
   0 & 1
 \end{array} \right)\left(\begin{array}{cc}
  1/D & 0 \\
   0 & D
 \end{array} \right)\left(\begin{array}{cc}
  1 & 0 \\
   C/D & 1
 \end{array} \right) ~.
\end{align}
Doing transformations in these two different ways will give insertions in different points and with different line integrals. The amplitudes should however match, and for the matching it will be important to consider the field independent factors that are produced like e.g. in \eqref{eq:transcase2withconstants}.


\begin{thebibliography}{99}


\bibitem{Bouwknegt:1992wg}
  P.~Bouwknegt and K.~Schoutens,
  ``$W$ symmetry in conformal field theory,''
  Phys.\ Rept.\  {\bf 223}, 183 (1993)
  [hep-th/9210010].

\bibitem{Henneaux:2010xg}
  M.~Henneaux and S.-J.~Rey,
  ``Nonlinear $W_{\infty}$ as asymptotic symmetry of three-dimensional higher spin anti-de Sitter gravity,''
  JHEP {\bf 1012}, 007 (2010)
  [arXiv:1008.4579 [hep-th]].

\bibitem{Campoleoni:2010zq}
  A.~Campoleoni, S.~Fredenhagen, S.~Pfenninger and S.~Theisen,
  ``Asymptotic symmetries of three-dimensional gravity coupled to higher-spin fields,''
  JHEP {\bf 1011}, 007 (2010)
  [arXiv:1008.4744 [hep-th]].


\bibitem{Gaberdiel:2010pz}
  M.~R.~Gaberdiel and R.~Gopakumar,
  ``An AdS$_3$ dual for minimal model CFTs,''
  Phys.\ Rev.\ D {\bf 83}, 066007 (2011)
  [arXiv:1011.2986 [hep-th]].

\bibitem{Prokushkin:1998bq}
  S.~F.~Prokushkin and M.~A.~Vasiliev,
  ``Higher spin gauge interactions for massive matter fields in 3-D AdS space-time,''
  Nucl.\ Phys.\ B {\bf 545}, 385 (1999)
  [hep-th/9806236].


\bibitem{Creutzig:2011fe}
  T.~Creutzig, Y.~Hikida and P.~B.~R\o nne,
  ``Higher spin AdS$_3$ supergravity and its dual CFT,''
  JHEP {\bf 1202}, 109 (2012)
  [arXiv:1111.2139 [hep-th]].

\bibitem{Candu:2012jq}
  C.~Candu and M.~R.~Gaberdiel,
  ``Supersymmetric holography on $AdS_3$,''
  JHEP {\bf 1309}, 071 (2013)
  [arXiv:1203.1939 [hep-th]].

\bibitem{Henneaux:2012ny}
  M.~Henneaux, G.~Lucena Gomez, J.~Park and S.-J.~Rey,
  ``Super-$W_\infty$ asymptotic symmetry of higher-spin $AdS_3$ supergravity,''
  JHEP {\bf 1206}, 037 (2012)
  [arXiv:1203.5152 [hep-th]].


\bibitem{Creutzig:2012ar}
  T.~Creutzig, Y.~Hikida and P.~B.~R\o nne,
  "$\mathcal{N}=1$ supersymmetric higher spin holography on AdS$_3$,''
  JHEP {\bf 1302} (2013) 019
  [arXiv:1209.5404 [hep-th]].

\bibitem{Beccaria:2013wqa}
  M.~Beccaria, C.~Candu, M.~R.~Gaberdiel and M.~Groher,
  ``$\mathcal{N}=1$ extension of minimal model holography,''
  JHEP {\bf 1307}, 174 (2013)
  [arXiv:1305.1048 [hep-th]].

\bibitem{Gaberdiel:2013vva}
  M.~R.~Gaberdiel and R.~Gopakumar,
  ``Large $\mathcal{N}=4$ Holography,''
  JHEP {\bf 1309}, 036 (2013)
  [arXiv:1305.4181 [hep-th]].

\bibitem{Creutzig:2013tja}
  T.~Creutzig, Y.~Hikida and P.~B.~R\o nne,
  ``Extended higher spin holography and Grassmannian models,''
  JHEP {\bf 1311}, 038 (2013)
  [arXiv:1306.0466 [hep-th]].

\bibitem{Creutzig:2014ula}
  T.~Creutzig, Y.~Hikida and P.~B.~R\o nne,
  ``Higher spin AdS$_{3}$ holography with extended supersymmetry,''
  JHEP {\bf 1410}, 163 (2014)
  [arXiv:1406.1521 [hep-th]].

\bibitem{Gaberdiel:2014cha}
  M.~R.~Gaberdiel and R.~Gopakumar,
  ``Higher Spins \& Strings,''
  JHEP {\bf 1411}, 044 (2014)
  [arXiv:1406.6103 [hep-th]].

\bibitem{Alday:2009aq}
  L.~F.~Alday, D.~Gaiotto and Y.~Tachikawa,
  ``Liouville correlation functions from four-dimensional gauge theories,''
  Lett.\ Math.\ Phys.\  {\bf 91}, 167 (2010)
  [arXiv:0906.3219 [hep-th]].

\bibitem{Wyllard:2009hg}
  N.~Wyllard,
  ``$A_{N-1}$ conformal Toda field theory correlation functions from conformal ${\cal N} = 2$ SU$(N)$ quiver gauge theories,''
  JHEP {\bf 0911}, 002 (2009)
  [arXiv:0907.2189 [hep-th]].



\bibitem{Ribault:2005wp}
  S.~Ribault and J.~Teschner,
  ``$H^+_3$-WZNW correlators from Liouville theory,''
  JHEP {\bf 0506}, 014 (2005)
  [hep-th/0502048].

\bibitem{HS}
  Y.~Hikida and V.~Schomerus,
  ``$H^+_3$ WZNW model from Liouville field theory,''
  JHEP {\bf 0710}, 064 (2007)
  [arXiv:0706.1030 [hep-th]].

\bibitem{HS2}
  Y.~Hikida and V.~Schomerus,
  ``Structure constants of the OSP(1$|$2) WZNW model,''
  JHEP {\bf 0712}, 100 (2007)
  [arXiv:0711.0338 [hep-th]].

\bibitem{CHR}
  T.~Creutzig, Y.~Hikida and P.~B.~R\o nne,
  ``Supergroup - extended super Liouville correspondence,''
  JHEP {\bf 1106}, 063 (2011)
  [arXiv:1103.5753 [hep-th]].

\bibitem{K}
  V.~G.~Knizhnik,
  ``Superconformal algebras in two-dimensions,''
  Theor.\ Math.\ Phys.\  {\bf 66}, 68 (1986)
  [Teor.\ Mat.\ Fiz.\  {\bf 66}, 102 (1986)].
\bibitem{B}
  M.~A.~Bershadsky,
  ``Superconformal algebras in two-dimentions with arbitrary $N$,''
  Phys.\ Lett.\  B {\bf 174}, 285 (1986).

\bibitem{KW}
  V.~G.~Kac and M.~Wakimoto,
  ``Quantum reduction and representation theory of superconformal algebras,''
  math-ph/0304011.

\bibitem{deBoer:1993iz}
  J.~de Boer and T.~Tjin,
  ``The Relation between quantum $W$ algebras and Lie algebras,''
  Commun.\ Math.\ Phys.\  {\bf 160}, 317 (1994)
  [hep-th/9302006].

\bibitem{Polyakov}
  A.~M.~Polyakov,
  ``Gauge transformations and diffeomorphisms,''
  Int.\ J.\ Mod.\ Phys.\  A {\bf 5}, 833 (1990).


\bibitem{Bershadsky}
  M.~Bershadsky,
  ``Conformal field theories via Hamiltonian reduction,''
  Commun.\ Math.\ Phys.\  {\bf 139}, 71 (1991).

\bibitem{Bais:1990bs}
  F.~A.~Bais, T.~Tjin and P.~van Driel,
  ``Covariantly coupled chiral algebras,''
  Nucl.\ Phys.\  B {\bf 357}, 632 (1991).


\bibitem{Zamolodchikov:1985wn}
  A.~B.~Zamolodchikov,
  ``Infinite additional symmetries in two-dimensional conformal quantum field theory,''
  Theor.\ Math.\ Phys.\  {\bf 65}, 1205 (1985)
  [Teor.\ Mat.\ Fiz.\  {\bf 65}, 347 (1985)].


\bibitem{Romans}
  L.~J.~Romans,
  ``Quasisuperconformal algebras in two-dimensions and Hamiltonian reduction,''
  Nucl.\ Phys.\  B {\bf 357}, 549 (1991).

\bibitem{Creutzig:2010zp}
  T.~Creutzig and Y.~Hikida,
  ``Branes in the OSP$(1|2)$ WZNW model,''
  Nucl.\ Phys.\ B {\bf 842}, 172 (2011)
  [arXiv:1004.1977 [hep-th]].

\bibitem{FZZ}
 V.~A.~Fateev, A.~B.~Zamolodchikov and Al.~B.~Zamolodchikov,
unpublished.


\bibitem{HS3}
  Y.~Hikida and V.~Schomerus,
  ``The FZZ-duality conjecture - A proof,''
  JHEP {\bf 0903}, 095 (2009)
  [arXiv:0805.3931 [hep-th]].


\bibitem{Creutzig:2010bt}
  T.~Creutzig, Y.~Hikida and P.~B.~R\o nne,
  ``The FZZ duality with boundary,''
  JHEP {\bf 1109}, 004 (2011)
  [arXiv:1012.4731 [hep-th]].

\bibitem{deBoer:1992sy}
  J.~de Boer and T.~Tjin,
  ``Quantization and representation theory of finite $W$ algebras,''
  Commun.\ Math.\ Phys.\  {\bf 158}, 485 (1993)
  [arXiv:hep-th/9211109].

\bibitem{Bars}
  I.~Bars,
  ``Free fields and new cosets of current algebras,''
  Phys.\ Lett.\  B {\bf 255}, 353 (1991).


\bibitem{Feigin:2004wb}
  B.~L.~Feigin and A.~M.~Semikhatov,
  ``$W^{(2)}_n$ algebras,''
  Nucl.\ Phys.\ B {\bf 698} (2004) 409
  
  
  \bibitem{Creutzig:2014lsa}
    T.~Creutzig and A.~R.~Linshaw,
    ``Cosets of affine vertex algebras inside larger structures,''
    arXiv:1407.8512 [math.RT].



\bibitem{Alday:2010vg}
  L.~F.~Alday and Y.~Tachikawa,
  ``Affine SL$(2)$ conformal blocks from 4d gauge theories,''
  Lett.\ Math.\ Phys.\  {\bf 94}, 87 (2010)
  [arXiv:1005.4469 [hep-th]].

\bibitem{Kozcaz:2010yp}
  C.~Kozcaz, S.~Pasquetti, F.~Passerini and N.~Wyllard,
  ``Affine sl$(N)$ conformal blocks from $\mathcal{N}=2$ SU$(N)$ gauge theories,''
  JHEP {\bf 1101}, 045 (2011)
  [arXiv:1008.1412 [hep-th]].

  \bibitem{Frenkel:2015rda}
    E.~Frenkel, S.~Gukov and J.~Teschner,
    ``Surface operators and separation of variables,''
    arXiv:1506.07508 [hep-th].

\bibitem{Wyllard:2010rp}
  N.~Wyllard,
  ``$W$-algebras and surface operators in $\mathcal{N}=2$ gauge theories,''
  J.\ Phys.\ A {\bf 44}, 155401 (2011)
  [arXiv:1011.0289 [hep-th]].

\bibitem{Wyllard:2010vi}
  N.~Wyllard,
  ``Instanton partition functions in $\mathcal{N}=2$ SU$(N)$ gauge theories with a general surface operator, and their $W$-algebra duals,''
  JHEP {\bf 1102}, 114 (2011)
  [arXiv:1012.1355 [hep-th]].
    

\bibitem{Fay}
  J.~Fay,
  ``Theta functions on Riemann surfaces,''
  Lecture Notes in Mathematics 352,
  Springer-Verlag (1973).

\bibitem{Mumford}
  D.~Mumford,
  ``Tata lectures on theta, Vols. I, II,''
  Progress in Mathematics 43,
  Birkh\"auser (1984)

\bibitem{AMV}
  L.~Alvarez-Gaume, G.~W.~Moore and C.~Vafa,
  ``Theta functions, modular invariance, and strings,''
  Commun.\ Math.\ Phys.\  {\bf 106}, 1 (1986).

\bibitem{VV}
  E.~P.~Verlinde and H.~L.~Verlinde,
  ``Chiral bosonization, determinants and the string partition function,''
  Nucl.\ Phys.\  B {\bf 288}, 357 (1987).

\bibitem{Bernard}
  D.~Bernard,
  ``On the Wess-Zumino-Witten models on Riemann surfaces,''
  Nucl.\ Phys.\  B {\bf 309}, 145 (1988).


\bibitem{Tjin:1991wm} 
  T.~Tjin and P.~Van Driel,
  ``Coupled WZNW - Toda models and covariant kdV hierarchies,''
  ITFA-91-04.




\end{thebibliography}
\end{document}